\documentclass[aps,pra,superscriptaddress,showpacs,showkeys]{revtex4}
\usepackage{epsfig}
\usepackage{graphicx}
\usepackage{bm}
\usepackage{amssymb}
\usepackage{amsmath}
\newcommand{\zh}{\bm}

\newcommand{\mee}{{E}}

\newcommand{\dee}{{\varepsilon}}

\newcommand{\zhr}{{\zh r}}

\newcommand{\zhe}{{\zh e}}

\newcommand{\zhp}{{\zh p}}
\newcommand{\zhk}{{\zh k}}
\newcommand{\zhalpha}{{\zh\alpha}}

\newcommand{\zhnu}{{\zh\nu}}

\newcommand{\zhsigma}{{\zh\sigma}}

\newcommand{\hH}{{\hat{H}}}
\newcommand{\hXi}{{\Delta\hat{V}}}
\newcommand{\setg}{{g}}
\newcommand{\bpsi}{{\bar{\psi}}}

\newcommand{\Br}[1]{(\ref{#1})}
\newcommand{\Eq}[1]{Eq.\ (\ref{#1})}

\newcommand{\Fig}[1]{Fig.\ \ref{#1}}
\newcommand{\txt}[1]{{\rm #1}}

\newcommand{\vext}{V^{\txt{ext}}}

\newcommand{\vectorb}[2]{
\left(
\begin{array}{c}
#1\\
#2\\
\end{array}
\right)
        }

\begin{document}

\title{QED theory of elastic electron scattering on hydrogen-like ions involving formation and decay of autoionizing states}

\author{ K.\ N.\ Lyashchenko}
 \affiliation{Department of Physics,
             St.\ Petersburg State University, 7/9 Universitetskaya nab.,
             St. Petersburg, 199034, Russia}
\affiliation{Institute of Modern Physics, Chinese Academy of Sciences, Lanzhou 730000, China}
\author{ D.\ M.\ Vasileva}
\affiliation{Department of Physics,
             St.\ Petersburg State University, 7/9 Universitetskaya nab.,
             St. Petersburg, 199034, Russia}
\author{O.\ Yu.\ Andreev}
\email{o.y.andreev@spbu.ru}
\affiliation{Department of Physics,
             St.\ Petersburg State University, 7/9 Universitetskaya nab.,
             St. Petersburg, 199034, Russia}
\affiliation{Center for Advanced Studies, Peter the Great St. Petersburg Polytechnic University, 195251 St. Petersburg, Russia}
\author{ A. B. Voitkiv }
\affiliation{ Institute for Theoretical Physics I,
Heinrich-Heine-University of D\"usseldorf, D\"usseldorf 40225, Germany }

\date{\today}

\begin{abstract}
We develop {\it ab initio} relativistic QED theory for elastic electron scattering on hydrogen-like highly charged ions for impact energies where, 
in addition to direct (Coulomb) scattering, the process can also proceed via formation and consequent Auger decay of autoionizing states of the corresponding helium-like ions.
Even so the primary goal of the theory is to treat electron scattering on highly charged ions, a comparison with experiment shows that it can also be applied for 
relatively light ions covering thus a very broad range of the scattering systems.
Using the theory we performed calculations for elastic electron scattering on B$^{4+}$, Ca$^{19+}$, Fe$^{25+}$, Kr$^{35+}$, 
and Xe$^{53+}$. The theory was also generalized for collisions of hydrogen-like highly charged ions with atoms 
considering the latter as a source of (quasi-) free electrons.
\end{abstract}

\maketitle

\section{Introduction}

Atomic systems with two or more electrons possess autoionizing states which can be highly visible 
in many processes studied by atomic physics, e.~g. photo and impact ionization, dielectronic recombination 
and photon or electron scattering. In particular, when an electron is incident on an ion, for certain (resonant) energies
of the incident electron an autoionizing state can be formed.
This state can then decay either via spontaneous radiative decay
or via Auger decay due to the electron-electron interaction.
In the former case dielectronic recombination takes place
whereas in the latter resonant electron scattering occurs.
Depending on whether the initial and final ionic states coincide
or not, the energy of the scattered electron can be equal
to the energy of the incident electron (elastic resonant scattering) or differ from it (inelastic resonant scattering).

An incident electron can also scatter on an ion
without excitation of the internal degrees of freedom of the ion.
In such a case scattering proceeds via the Coulomb force
acting between the electron and the (partially screened) nucleus of the
ion. This scattering channel is non-resonant and elastic
and its amplitude
should be added coherently to the amplitude for the elastic resonant
scattering. As a result, there appears interference between these two channels 
which has to be taken into account in a proper treatment of elastic scattering. 

Resonant scattering was extensively studied
for non-relativistic electrons incident on light ions (for instance,
$e^-$ + He$^+$(1s) $ \to $ He$^{**}$ $\to$ He$^+$(1s) + $e^-$).
This scattering becomes especially important at large
scattering angles (as viewed in the rest frame of the ion)
where the contribution of the potential Coulomb scattering is minimal.
Therefore, experimental investigations (which often replace electron-ion
collisions by ion-atom collisions in which atomic electrons are regarded
as quasi-free) were mainly focused on such angles (see e.g.
\cite{huber1994PhysRevLett.73.2301,greenwood1995PhysRevLett.75.1062,troth1996PhysRevA.54.R4613,zouros2003PhysRevA.68.010701,benis2004pra69-052718,benis2006PhysRevA.73.029901}).
There exists also a large number of calculations
for resonant electron-ion scattering in which
a non-relativistic electron interacts with a light ion
(see e.g. \cite{griffin1990PhysRevA.42.248,bhalla1990PhysRevLett.64.1103,benis2004pra69-052718,benis2006PhysRevA.73.029901}).

In sharp contrast, the studies on resonant
scattering of an electron on a highly charged ion, in which relativistic and QED effects can become of importance, 
are almost absent with no experimental data and merely one theoretical paper \cite{kollmar2000}
in which scattering of an electron on a hydrogen-like uranium ion was
considered. Moreover, in \cite{kollmar2000} just the resonant part
of the scattering was calculated whereas the potential Coulomb part
as well as the interference between them were not considered.

In case of scattering on highly charged ions the electrons are subjected to a very strong field generated 
by the ionic nucleus. As a result, the account of relativistic and QED effects may become of great importance for a proper description of the scattering process.

In the present paper we consider elastic electron scattering on hydrogen-like highly charged ions for impact energies 
where the presence of autoionizing states of the corresponding helium-like ions can actively influence this process. 
To this end we shall develop, for the first time, {\it ab initio} relativistic QED theory of this process, 
which enables one to address both its resonant and Coulomb parts in an unified and self-consistent way.

Relativistic units are used throughout unless otherwise is stated. 

\section{General theory}
\label{section-tp}
We consider elastic electron scattering on a hydrogen-like ion which is initially in its ground state, 
\begin{eqnarray}
e^-_{\zhp_i,\mu_i} +X^{(Z-1)+}(1s)
&\to&\label{eqn170705n01}
e^-_{\zhp_f,\mu_f}
+
X^{(Z-1)+}(1s)
\,,
\end{eqnarray}
where $Z$ is the atomic number of the ion $X$.  
If the energy of the initial state of the electron system, which consists of 
the incident electron $e^-_{\zhp,\mu}$ with an asymptotic momentum $\zhp$ and polarization $\mu$ 
and the $1s$-electron bound in the ion, 
is close to the energy of a doubly excited (autoionizing) state of the corresponding helium-like ion, 
the resonant scattering channel,  
\begin{eqnarray}
e^-_{\zhp_i,\mu_i}+X^{(Z-1)+}(1s)
&\to&\label{eqn170705n02}
X^{(Z-2)+}(d)
\,\to\,
e^-_{\zhp_f,\mu_f}+
X^{(Z-1)+}(1s)
\,,
\end{eqnarray}
where $d$ is a doubly excited state, 
becomes of importance.  
Here, the scattering proceeds via the formation of
a doubly excited state ($d$) and its subsequent Auger decay.
This channel is driven by the interelectron interaction.

One can expect that in case of highly charged ions 
the main channel of the electron-ion scattering
process \Br{eqn170705n01} is Coulomb scattering, which is non-resonant.
Therefore, in order to investigate the resonant structure experimentally 
the electron scattering to very large angles ( $\simeq 180^{\circ}$), 
for which the contribution of the main channel is minimal, should be 
considered \cite{benis2004pra69-052718}.
Accordingly, one needs to calculate 
the differential cross section.

We shall consider the scattering process 
\Br{eqn170705n01} using the Furry picture 
\cite{furry51}, in which the action of a strong external field
(for example, the field of the ionic nucleus) on the electrons 
is taken into account from the very beginning. 
The electrons interact with each other via the interaction
with the quantized electromagnetic and electron-positron fields, 
which is accounted for by using perturbation theory.

The in- $(+)$ and out-going $(-)$ wave functions of an electron
in an external central electric field
with a asymptotic momentum ($\zhp$) and polarization ($\mu$)
can be presented as \cite{akhiezer65b} 
\begin{eqnarray}
\psi^{(\pm)}_{\zhp\mu}(\zhr)
&=&\label{eqn181017n02}
N
\sum\limits_{jlm} \Omega^{+}_{jlm}(\zhnu)v_{\mu}(\zhnu)
e^{\pm i\phi_{jl}} i^{l}
\psi_{\varepsilon jlm}(\zhr)
\,,
\end{eqnarray}
where
\begin{eqnarray}
N
&=&
\frac{(2\pi)^{3/2}}{\sqrt{p\varepsilon}}
\end{eqnarray}
is the normalization factor,
$\zhnu=\zhp/|\zhp|$ is the unit vector 
defining the angular dependence of the momentum $\zhp$. 
The wave functions $\psi_{\varepsilon jlm}(\zhr)$ 
describe electrons with 
the energy $\varepsilon=\sqrt{1+\zhp^2}$,
the total angular momentum $j$, its projection $m$ 
and parity defined by the orbital momentum $l$, and 
$\phi_{jl}$ are the phases determined by the external field
(see \Eq{coulomb-phase}
in Appendix~\ref{appendixb}). 
These wave functions are normalized according to 
\begin{eqnarray}
\int d\zhr\,
\psi^{+}_{\varepsilon' j'l'm'}(\zhr)
\psi_{\varepsilon jlm}(\zhr)
&=&
\delta(\varepsilon'-\varepsilon)
\delta_{j'j}
\delta_{l'l}
\delta_{m'm}
\,,
\end{eqnarray}
where $\delta$ denotes either the delta function 
or the Kronecker symbol, respectively.
The spherical bispinor reads 
\cite{varshalovich}
\begin{eqnarray}
\Omega_{jlm}(\zhnu)
&=&\label{omega}
\sum\limits_{m_lm_s}C^{ls}_{jm}(m_lm_s) Y_{lm_l}(\zhnu)\chi_{m_s}
\,,
\end{eqnarray}
where $C^{ls}_{jm}(m_lm_s)=\langle lm_l sm_s|jm \rangle$
are the Clebsch-Gordan coefficients,
$Y_{lm_l}(\zhnu)$ are spherical harmonics and
$\chi_{m_s}$, 
\begin{eqnarray}
\chi_{+1/2}
&=&
\vectorb{1}{0}
\,,\qquad
\chi_{-1/2}
\,=\,
\vectorb{0}{1}
\,, 
\end{eqnarray}
are spinors. Further, in \Br{eqn181017n02} there are also spinors $v_{\mu}(\zhnu)$, which are defined according to 
\begin{eqnarray}
\frac{1}{2}\zhnu\zhsigma
v_{\mu}(\zhnu)
&=&
\mu
v_{\mu}(\zhnu)
\,,
\\
(v_{\mu'}(\zhnu),v_{\mu}(\zhnu))
&=&
\delta_{\mu'\mu}
\,,
\\
v_{\mu}(\zhe_z)
&=&
\chi_{\mu}
\,,
\end{eqnarray}
where $\zhsigma$ is the Pauli vector,
$\zhe_z$ is the unit vector along the z-axis.
Then the wave functions $\psi^{(\pm)}_{\zhp\mu}(\zhr)$
are normalized as
\begin{eqnarray}
\int d\zhr\,
\psi^{(\pm)+}_{\zhp'\mu'}(\zhr)
\psi^{(\pm)}_{\zhp\mu}(\zhr)
&=&\nonumber
(2\pi)^3
\delta(\zhp'-\zhp)
\delta_{\mu'\mu}
\\
&=&\label{norm-psipmu}
\frac{(2\pi)^3}{\varepsilon p}
\delta(\varepsilon'-\varepsilon)
\delta(\cos\theta'-\cos\theta)
\delta(\varphi'-\varphi)
\delta_{\mu'\mu}
\,.
\end{eqnarray}

For the process of elastic electron scattering
\Br{eqn170705n01}
the contribution of its Coulomb part to the amplitude is very important.
The Coulomb scattering amplitude is usually calculated by studying
the asymptotics of the electron wave functions in the Coulomb field
\cite{akhiezer65b}.

The resonant part of the scattering process 
\Br{eqn170705n02}
is due to the formation and decay of autoionizing states of the corresponding helium-like ion.  
In the present paper, for the description of autoionizing states 
within the QED theory
the line-profile approach (LPA) will be employed
\cite{andreev08pr}, where
the QED perturbation theory is used. 
In particular, it is necessary to take into account
various corrections such as the interelectron interaction corrections and
the relativistic corrections to the scattering amplitude.

In order to make the consideration of both Coulomb and resonant parts of the scattering amplitude
self-consistent, we shall apply
the formal theory of scattering for the Coulomb potential 
considering it by using perturbation theory.

\subsection{Coulomb scattering amplitude} 

For simplicity, in this subsection we limit ourselves 
to the consideration of a one-electron system.

Within the scattering theory developed in
\cite{gell-mann1953PhysRev.91.398,lippmann1950PhysRev.79.469}
the in- $(+)$ and out- $(-)$ states satisfy 
the Lippmann-Schwinger equation
\begin{eqnarray}
\psi^{(\pm)}_{\zhp\mu}(\zhr)
&=&\label{eqn181017n04}
\phi_{\zhp\mu}(\zhr)
+\frac{1}{\varepsilon-\hat{H}_0\pm i0}\hat{V}\psi^{(\pm)}_{\zhp\mu}(\zhr)
\,,
\end{eqnarray}
where $\hat{V}$ represents the scattering potential.
Here, the function $\phi_{\zhp\mu}(\zhr)$ describes a free electron and
can be obtained from \Eq{eqn181017n02} by setting $Z=0$
\begin{eqnarray}
\phi_{\zhp\mu}(\zhr)
&=&
u_{\mu}(p)
e^{i\zhp\zhr}
\,,
\\
u_{\mu}(p)
&=&
\frac{1}{\sqrt{2\varepsilon}}
\vectorb{\sqrt{\varepsilon+1}v_{\mu}(\zhnu)}{\sqrt{\varepsilon-1}(\zhsigma\zhnu)v_{\mu}(\zhnu)}
\,.
\end{eqnarray}
Following
\cite{gell-mann1953PhysRev.91.398}
we introduce the $R$-matrix elements according to 
\begin{eqnarray} 
R_{if}
&=&\nonumber
\langle
\phi^{}_{\zhp_f\mu_f}
|\hat{V}|
\psi^{(+)}_{\zhp_i\mu_i}
\rangle
\\
&=&\label{eqn181017n05} 
\langle
\phi^{}_{\zhp_f\mu_f}
|\hat{V}|
\phi_{\zhp_i\mu_i}
\rangle
+
\langle
\phi^{}_{\zhp_f\mu_f}
|\hat{V}
\frac{1}{\varepsilon_i-\hat{H}_0 + i0}\hat{V}
|
\psi^{(+)}_{\zhp_i\mu_i}
\rangle
\,.
\end{eqnarray}
The $S$-matrix elements is related to the $R$-matrix as follows 
\begin{eqnarray}
S_{if}
&=&\label{eqn181017n06}
\langle
\phi^{}_{\zhp_f\mu_f}
|\hat{S}|
\phi^{}_{\zhp_i\mu_i}
\rangle
\,=\,
\delta_{if}
-2\pi i \delta(\varepsilon_i-\varepsilon_f) R_{if}
\,.
\end{eqnarray}
Since the $S$-matrix elements can be also written as
\begin{eqnarray}
S_{if}
&=&\label{eqn181017n03}
\langle
\psi^{(-)}_{\zhp_f\mu_f}
|
\psi^{(+)}_{\zhp_i\mu_i}
\rangle
\, 
\end{eqnarray}
we obtain that ($f \neq i$) 
\begin{eqnarray}
\langle \psi^{(-)}_{\zhp_f\mu_f} | \psi^{(+)}_{\zhp_i\mu_i} \rangle
&=&\label{eqn181026n001}
(-2\pi i)\delta(\varepsilon_f-\varepsilon_i)
\,
R_{if}
\,. 
\end{eqnarray}
In the case of Coulomb scattering, 
$\hat{V} = - \alpha Z/r $, the states 
$ \psi^{(\pm)}_{\zhp_i\mu_i} $ 
can be conveniently evaluated using 
the expansions \Br{eqn181017n02}, which enables us 
to obtain the following expression 
\begin{eqnarray}
R^\txt{Coul}_{if}
&=&\label{eqn181017n01}
N^2
\frac{(-1)p}{(2\pi)^2}
\sum_{m}
(v_{\mu_f}(\zhnu_f))_m^{*}
M_{m\mu_i}(\theta,\varphi)
\, 
\end{eqnarray}
(a detailed derivation of $R^\txt{Coul}$ is presented in
Appendix \ref{appendixa}). 
Following \cite{burke2011b}
we introduced the matrix $M$
\begin{align}
M_{+1/2,+1/2}(\theta,\varphi)&=f(\theta),&M_{+1/2,-1/2}(\theta,\varphi)&=g(\theta)e^{-i\varphi}\,,
\nonumber\\
M_{-1/2,+1/2}(\theta,\varphi)&=-g(\theta)e^{i\varphi},&M_{-1/2,-1/2}(\theta,\varphi)&=f(\theta)
\,,
\end{align}
where $f(\theta)$ and $g(\theta)$ are the relativistic scattering amplitudes
\begin{eqnarray}
f(\theta)
&=&
\dfrac{1}{2\pi i}
\sum_{jl}|\varkappa|(e^{2i\phi_\varkappa}-1)P_l(\cos\theta)\\
g(\theta)
&=&
\dfrac{1}{2\pi i}
\sum_{l}(e^{2i\phi_{\varkappa=-l-1}}-e^{2i\phi_{\varkappa=l}})P^1_l(\cos\theta)
\,.
\end{eqnarray}

\subsection{Coulomb scattering amplitude within the LPA}
In the QED theory
the scattering matrix ($S$) can be represented as the sum of the normal ordered products
of field operators corresponding to different processes of particle scattering.
Each normal ordered product and, consequently, any scattering process
can be represented graphically according to the Feynman rules.
In the corresponding matrix elements the integration over
$x^{\mu}=(t,\zhr)$ 4-vectors is performed.
If in the process the energy is conserved, 
after the integration over the time variables 
the matrix element can be written as 
\cite{akhiezer65b}
\begin{eqnarray}
S^\txt{QED}_{if}
&=&\label{eqn181026n08}
(-2\pi i)\delta(\varepsilon_f-\varepsilon_i) U^\txt{QED}_{if}
\,,
\end{eqnarray}
where $U^\txt{QED}_{if}$ is the amplitude.
In QED of strong fields
the $S$-matrix elements and, correspondingly, the amplitude $U_{if}$
are evaluated with the use of perturbation theory
\cite{akhiezer65b}.
For the description of highly charged ions within QED theory  
special methods are employed.
Most of QED calculations were performed using the adiabatic S-matrix approach
\cite{gellmann51,sucher57,labzowsky70},
the two-time Green's function method
\cite{shabaev02},
the covariant-evolution-operator method
\cite{lindgren04}
and the LPA
\cite{andreev08pr}.
In the present paper we employ the LPA.

We shall now establish the relationship between the amplitude 
$R^\txt{Coul}_{if}$ defined by
\Eq{eqn181017n05}
and the corresponding amplitudes derived within the LPA. 

Let us evaluate the scattering amplitude corresponding to the interaction with an external field $V^\txt{ext}$ in the one-electron case. 
Within the LPA the initial and final states (the reference states) are described as resonances in the process of scattering.
As an auxiliary process it is normally most convenient to take 
elastic photon scattering.
For the properties of the reference states to be independent of 
the details of the scattering process,
the resonance approximation is employed \cite{andreev08pr}.
In this approximation the line profile is interpolated by the Lorentz contour, the position of the resonance and its width define 
the energy and width of the corresponding state.

\begin{figure}[h]
\includegraphics[width=8pc]{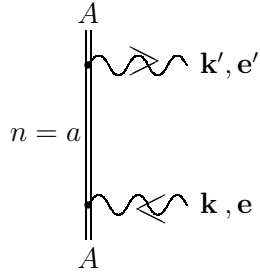}
\caption{
Feynman graph describing the elastic photon scattering on an atomic electron.
The double solid line denotes the electron in the external potential $V^\txt{F}$
(the Furry picture).
The wavy lines with the arrows describe the emission and absorption of
a photon with momentum $\zhk$ and polarization $e$.
}
\label{fig-06}
\end{figure}

We consider the process of elastic photon scattering on one-electron ion initially being in its ground state.
The Feynman graphs corresponding to this process in the zeroth order of the perturbation theory are depicted in \Fig{fig-06}.
%
%
%
%
The corresponding $S$-matrix element reads
\cite{akhiezer65b,andreev08pr}
\begin{eqnarray}
S^\txt{QED}_{if}
&=&\label{eqn181026n02x}
(-ie)^2
\int d^4 x_u d^4 x_d\,
\bpsi_{u}(x_u)
\gamma^{\mu_u}
S(x_u,x_d)
\gamma^{\mu_d}
\psi_{d}(x_d)
A^{*(k',\lambda')}_{\mu_u}(x_u)
A^{(k,\lambda)}_{\mu_d}(x_d)
\,.
\end{eqnarray}
We assume that $\psi_u=\psi_d$ describes the $1s$-electron.
The electron propagator $S(x_u,x_d)$ can be written as
\begin{eqnarray}
S(x_u,x_d)
&=&
\frac{i}{2\pi}
\int d\omega_1 \,e^{-i\omega_1(t_u-t_d)}
\sum\limits_n
\frac{\psi_n(\zhr_u)\bpsi_n(\zhr_d)}{\omega_1-\dee_n(1-i0)}
\,,
\label{electr_propagator} 
\end{eqnarray}
where the sum runs over the entire Dirac spectrum.
Further, $A^{(k,\lambda)}_{\mu_d}(x_d)$ 
and $A^{*(k',\lambda')}_{\mu_u}(x_u)$ 
refer to the absorbed and emitted photons, respectively, 
$k^{\mu}=(\omega,\zhk)$ is the 4-vector of photon momentum,
$\lambda$ describes the photon polarization.
By inserting (\ref{electr_propagator}) into
\Eq{eqn181026n02x} 
and integrating over the time and frequency variables 
$t_u$, $t_d$ and $\omega_n$ we obtain 
\begin{eqnarray}
S^\txt{QED}_{if}
&=&\label{eqn181026n03}
(-2\pi i)\delta(\omega'+\dee_u-\omega-\dee_d)\,
e^2
\sum\limits_n
\frac{A^{*(k',\lambda')}_{un}A^{(k,\lambda)}_{nd}}{\omega+\dee_d-\dee_n}
\,,
\end{eqnarray}
where one-electron matrix elements 
\begin{eqnarray}
A^{(k,\lambda)}_{nd}
&=&
\int d^3 \zhr\,
\bpsi_{n}(\zhr)
\gamma^{\mu}
A^{(k,\lambda)}_{\mu}(\zhr)
\psi_d(\zhr)
\,,
\\
A^{*(k',\lambda')}_{un}
&=&
\int d^3 \zhr\,
\bpsi_{u}(\zhr)
\gamma^{\mu}
A^{*(k',\lambda')}_{\mu}(\zhr)
\psi_n(\zhr)
\,
\end{eqnarray}
were introduced. 

\begin{figure}[h]
\includegraphics[width=6pc]{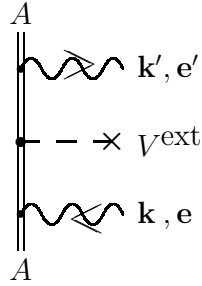}
\caption{
Feynman graph corresponding to the insertion describing the interaction with the external field
$V^\txt{ext}$.
The dotted line with a cross denotes the interaction with the field $V^\txt{ext}$. The notations are the same as in
\Fig{fig-06}.
}
\label{fig-06-ext}
\end{figure}

In the first order of the perturbation theory 
we consider one interaction with an external field
$A^\txt{ext}_{\mu}(x)=(V^\txt{ext}(\zhr),\zh{0})$.
The Feynman graph describing this interaction is presented in
\Fig{fig-06-ext}.
%
%
%
%
The corresponding $S$-matrix element reads
\begin{eqnarray}
S^{\txt{QED}(1)}_{if}
&=&\label{eqn181026n04-02}
(-ie)^3
\int d^4 x_u d^4 x_d d^4 x\,
\bpsi_{u}(x_u)
\gamma^{\mu_u}
A^{*(k',\lambda')}_{\mu_u}(x_u)
S(x_u,x)
\gamma^{\mu}
A^{\txt{ext}}_{\mu}(x)
S(x,x_d)
A^{(k,\lambda)}_{\mu_d}(x_d)
\gamma^{\mu_d}
\psi_{d}(x_d)
\,.
\end{eqnarray}
The integration over the time and frequency variables leads to
\begin{eqnarray}
S^{\txt{QED}(1)}_{if}
&=&
(-2\pi i)\delta(\omega'+\dee_u-\omega-\dee_d)\,
e^2
\sum\limits_{n_1 n_2}
\frac{A^{*(k',\lambda')}_{un_1}}{(\omega'+\dee_u-\dee_{n_1})}
\vext_{n_1 n_2}
\frac{A^{(k,\lambda)}_{n_2d}}
{(\omega+\dee_d-\dee_{n_2})}
\,,
\end{eqnarray}
where
\begin{eqnarray}
\vext_{n_1n_2}
&=&
e \int d^3 \zhr\,
\psi^{+}_{n_1}(\zhr)
\vext(\zhr)
\psi_{n_2}(\zhr)
\,.
\end{eqnarray}

Within the LPA the reference states 
(the initial and final states) 
are defined via resonances in 
the process of elastic photon scattering.
Accordingly, we are interested in 
(we note that $\dee_{u}=\dee_{d}$)
\begin{eqnarray}
\omega'+\dee_{u}
&\approx&
\dee_{f}
\,,
\\
\omega+\dee_{d}
&\approx&
\dee_{i}
\,.
\end{eqnarray}
In the resonance approximation, the reference states are described 
in such a way that their properties do not depend on a specific scattering process. Accordingly, retaining only terms with
${n_1}=f$, ${n_2}=i$,  
in the resonance approximation we get
\begin{eqnarray}
S^{\txt{QED}(1)\txt{res}}_{if}
&=&
(-2\pi i)\delta(\omega'+\dee_u-\omega-\dee_d)\,
e^2
\frac{A^{*(k',\lambda')}_{uf}}{(\omega'+\dee_u-\dee_{f})}
\vext_{fi}
\frac{A^{(k,\lambda)}_{id}}
{(\omega+\dee_d-\dee_{i})}
\,.
\end{eqnarray} 

The second order of the perturbation theory with respect to the interaction with the external field reads
\begin{eqnarray}
S^{\txt{QED}(2)}_{if}
&=&
(-2\pi i)\delta(\omega'+\dee_u-\omega-\dee_d)\,
e^2
\sum\limits_{n_1 n_2 n_3}
\frac{A^{*(k',\lambda')}_{un_1}}{(\omega'+\dee_u-\dee_{n_1})}
\vext_{n_1 n_2}
\frac{1}
{(\omega+\dee_{d}-\dee_{n_2})}
\vext_{n_2 n_3}
\frac{A^{(k,\lambda)}_{n_3d}}
{(\omega+\dee_{d}-\dee_{n_3})}
\,.
\end{eqnarray}
In the resonance approximation we keep only 
the term with $n_2 = n_3 = i$
\begin{eqnarray}
S^{\txt{QED}(2)\txt{res}}_{if}
&=&
(-2\pi i)\delta(\omega'+\dee_u-\omega-\dee_d)\,
e^2
\frac{A^{*(k',\lambda')}_{uf}}{(\omega'+\dee_u-\dee_{f})}
\vext_{f i}
\frac{1}
{(\omega+\dee_{d}-\dee_{i})}
\vext_{i i}
\frac{A^{(k,\lambda)}_{id}}
{(\omega+\dee_{d}-\dee_{i})}
\,.
\end{eqnarray}

The third order of perturbation theory 
in the resonance approximation reads
\begin{eqnarray}
S^{\txt{QED}(3)\txt{res}}_{if}
&=&\nonumber
(-2\pi i)\delta(\omega'+\dee_u-\omega-\dee_d)\,
\\
&&
\times
e^2
\sum\limits_{n_1 n_2}
\frac{A^{*(k',\lambda')}_{uf}}{(\omega'+\dee_u-\dee_{f})}
\vext_{f i}
\frac{1}
{(\omega+\dee_{d}-\dee_{i})}
\vext_{ii}
\frac{1}
{(\omega+\dee_{d}-\dee_{i})}
\vext_{i i}
\frac{A^{(k,\lambda)}_{id}}
{(\omega+\dee_{d}-\dee_{i})}
\,.
\end{eqnarray}
In the resonance approximation
the series of the perturbation theory 
composes a geometric progression, 
\begin{eqnarray}
\sum\limits_{l=0}^{\infty}
S^{\txt{QED}(l)\txt{res}}_{if}
&=&\nonumber
(-2\pi i)\delta(\omega'+\dee_u-\omega-\dee_d)\,
e^2
\sum\limits_{l=0}^{\infty}
\frac{A^{*(k',\lambda')}_{uf}}{(\omega'+\dee_u-\dee_{f})}
\vext_{f i}
\left[
\frac{1}{(\omega+\dee_{d}-\dee_{i})}
\vext_{i i}
\right]^l
\frac{A^{(k,\lambda)}_{id}}{(\omega+\dee_{d}-\dee_{i})}
\\
&=&
(-2\pi i)\delta(\omega'+\dee_u-\omega-\dee_d)\,
e^2
\frac{A^{*(k',\lambda')}_{uf}}{(\omega'+\dee_u-\dee_{f})}
\vext_{f i}
\frac{A^{(k,\lambda)}_{id}}{(\omega+\dee_{d}-\dee_{i}-\vext_{i i})}
\,.
\end{eqnarray}
Thus, the summation of the infinite series resulted 
in a shift of the position of the resonance, which 
represents a correction to the energy of the reference state.
This procedure was first applied in \cite{low52}
for the derivation of the natural line shape 
within the QED theory, where the correction due to 
the electron self-energy was considered.
We also note that the corrections 
to the photon scattering amplitude 
for a few electron ions were discussed 
in \cite{andreev08pr}. 

Within the LPA we introduce the amplitude
\begin{eqnarray}
U_{if}
&=&
\sum\limits^{\infty}_{l=1}
U^{(l)}_{if}
\,,
\label{amplitude}
\end{eqnarray}
where
\begin{eqnarray}
U^{(1)}_{if}
&=&
\vext_{f i}
\,,
\\
U^{(2)}_{if}
&=&
\sum\limits_{n}
\vext_{f n}
\frac{1}
{(\omega+\dee_{d}-\dee_{n})}
\vext_{n i}
\,,
\\
U^{(3)}_{if}
&=&
\sum\limits_{n_1 n_2}
\vext_{f n_1}
\frac{1}
{(\omega+\dee_{d}-\dee_{n_1})}
\vext_{n_1 n_2}
\frac{1}
{(\omega+\dee_{d}-\dee_{n_2})}
\vext_{n_2 i}
\,.
\label{amplitudes}
\end{eqnarray}
In the resonance approximation we can set $\omega+\dee_{d}=\dee_{i}$.
The amplitude (\ref{amplitude}) describes 
the transition ($i\to f$) caused by the external field $\vext$.

Introducing the operator $\hXi$, which is defined 
by its matrix elements as
\begin{eqnarray}
\Delta V_{n_1n_2}
&=&\label{eqn181026n07}
\vext_{n_1 n_2}
\end{eqnarray}
the amplitude can be rewritten as
\begin{eqnarray}
U_{if}
&=&\nonumber
\langle
\Psi^{(0)}_f|\hXi|\Psi^{(0)}_i
\rangle
+
\langle
\Psi^{(0)}_f|\hXi
\frac{1}{\varepsilon_i-\hH_0+i0}\hXi
|\Psi^{(0)}_i
\rangle
\\
&&\nonumber
+
\langle
\Psi^{(0)}_f|\hXi
\frac{1}{\varepsilon_i-\hH_0+i0}\hXi
\frac{1}{\varepsilon_i-\hH_0+i0}\hXi
|\Psi^{(0)}_i
\rangle
+\ldots
\\
&=&\label{eqn181019n01}
\sum\limits^{\infty}_{l=0}
\langle
\Psi^{(0)}_f|\hXi
\left[\frac{1}{\varepsilon_i-\hH_0+i0}\hXi\right]^l
|\Psi^{(0)}_i
\rangle
\,. 
\end{eqnarray}
The one-electron wave functions $\Psi^{(0)}_n$ 
describe electrons noninteracting with
the external field $\vext_{n_1 n_2}$
\begin{eqnarray}
\hH_0
\Psi^{(0)}_n
&=&
\varepsilon_n
\Psi^{(0)}_n
\,. 
\end{eqnarray}
Here, $\hH_0$ is the Dirac hamiltonian
\begin{eqnarray}
\hH_0
&=&
\zhalpha\hat{\zhp} + \beta m + V^\txt{F}(\zhr)
\,, 
\end{eqnarray}
where $\zhalpha$ and $\beta$ are the Dirac matrices 
and the choice of the potential $V^\txt{F}$ 
defines the Furry picture employed. 

The amplitude $U_{if}$ in
\Eq{eqn181019n01} can be also written as
\begin{eqnarray}
U_{if}
&=&\label{eqn181026n06}
\langle
\Psi^{(0)}_f|\hXi|\Phi_i
\rangle
\,,
\end{eqnarray}
where
\begin{eqnarray}
\Phi_i
&=&
\Psi^{(0)}_i
+
\frac{1}{\varepsilon_i-\hH_0+i0}\hXi
\Phi_i
\,. 
\label{120619_n1} 
\end{eqnarray}
The function $\Phi_i$
is a solution of the following equation
\begin{eqnarray}
(\hH_0+\hXi)
\Phi_i
&=&
\varepsilon_i
\Phi_i
\,.
\end{eqnarray}

We note that the amplitude (\ref{eqn181026n06}), 
in which the exact state is given by (\ref{120619_n1}), 
essentially coincides with the amplitude (\ref{eqn181017n05}), 
where the exact state is given by (\ref{eqn181017n04}).  
Therefore, we conclude that 
the amplitude $U_{if}$ obtained within the LPA -- a QED approach --
coincides with the amplitude $R_{if}$, which follows from (relativistic)  quantum mechanics and which, in particular, 
can be evaluated using \Eq{eqn181026n001}. 
One should add that in the framework of the LPA also QED corrections -- such as electron self-energy, vacuum polarization, photon exchange corrections -- 
can be taken into account using the procedure described above. 
Thus, in general, the operator $\Delta\hat{V}$ includes also 
the corresponding QED corrections.

\subsection{ Implementation of the LPA for the description 
of elastic resonant scattering } 

\begin{figure}[h]
\includegraphics[width=4pc]{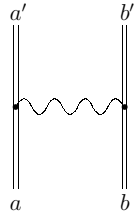}
\caption{
Feynman graph corresponding to the first-order interelectron
interaction in two-electron ion (one-photon exchange graph).
}
\label{fig-opx}
\end{figure}

\begin{figure}[h]
\includegraphics[width=13pc]{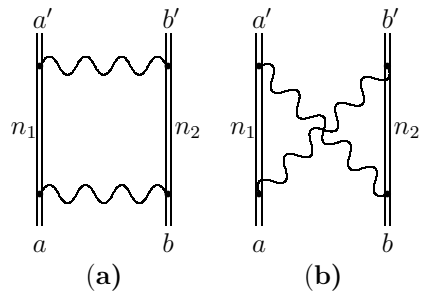}
\caption{
Feynman graphs corresponding to the second-order interelectron
interaction in two-electron ion (two-photon exchange graphs).
}
\label{fig-tpx}
\end{figure}

The scattering amplitude corresponding to the interelectron interaction
is given by Feynman graphs depicted in \Fig{fig-opx} and
\Fig{fig-tpx}. 
%
%
%
%
%
%
The graph in \Fig{fig-opx}
represents the one-photon exchange correction,
the graphs (a) and (b) in
\Fig{fig-tpx}
refer to the two-photon exchange corrections,   
the three- and more photon exchange corrections could also be considered.
One should note, however, that in the process under consideration 
all these graphs need a special treatment because they contain divergences. 

For the description of processes, which involve electrons 
interacting with the field of a highly charged nucleus, 
the Furry picture is normally used, in which the interaction 
of the electrons with the potential 
of the nucleus $V^\txt{F}= -\alpha Z/r$ is fully taken into account 
from the onset, whereas the interelectron interaction 
is considered using a QED perturbation theory. However, 
in the case of {\it elastic} scattering such 
an approach leads to divergent results arising from 
the application of the perturbation theory to a long-range Coulomb interaction
between bound and free electrons.  

In order to modify the `standard' approach we note that 
the incident and scattered electron most of the time 
is moving in the Coulomb field of a hydrogen-like ion 
with the net charge $Z-1$. Therefore, we shall avoid 
the use of the perturbation theory for the 
interaction of the free electron 
with the bound electron by employing the Furry picture 
in which the free electron is supposed to move in the field 
of the nuclear charge $Z-1$. Then the interaction of 
the free electron with the bound electron together
with its interaction with the `remaining' charge  
of the nucleus [$Z-(Z-1)=1$] is already 
a short-range interaction since it corresponds to the interaction 
with an electrically neutral system
[the charge of the bound electron ($e$) smeared out over 
the size of its bound state and the `remaining' charge 
of the nucleus ($-e$)]. Now the divergences do not arise and 
this short-range interaction can be accounted for by using 
the perturbation theory.

Thus, our consideration involves the following points. 
i) The wave functions of the incident 
and scattered electron are obtained by 
solving the Dirac equation with the potential $-\alpha (Z-1)/r$. 
ii) The wave functions of all other electrons (bound and virtual) 
are derived from the Dirac equation with the potential  
$-\alpha Z/r$. iii) The interaction of the continuum electrons 
with the `remaining' charge of the nucleus
is calculated with the use of perturbation theory. 
iv) The interelectron interaction is considered as the interaction 
with the quantized electromagnetic and electron-positron fields 
within the QED perturbation theory.
v) The divergences arising from the long-range Coulomb interaction 
of the continuum electron with
the `remaining' charge of the nucleus and with the bound electron
can be regularized and cancel each other.

\begin{figure}[h]
\includegraphics[width=20pc]{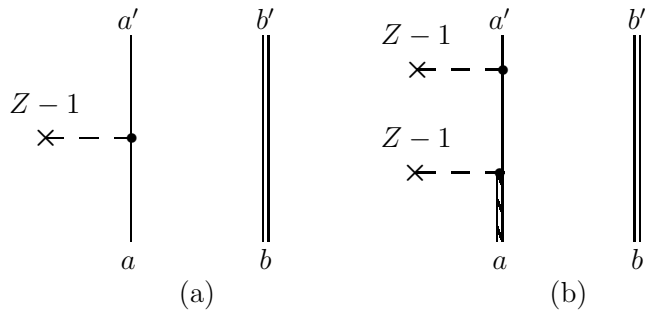}
\caption{
Feynman graphs representing the first (a) and  second (b) terms in the right side of
\Eq{eqn181017n05}
for two-electron ion.
The single solid line denotes free electron, the double line denotes
electron in the field of potential $V^\txt{F}= -\alpha Z/r$.
The dashed line with the cross designates interaction with external field
$V_\txt{cont} = -\alpha (Z-1)/r$.
The double striped line describes electron in the Furry picture with potential
$V^\txt{F}_\txt{cont} = -\alpha (Z-1)/r$.
}
\label{fig-coul-1}
\end{figure}

The amplitude of the Coulomb scattering given by
\Eq{eqn181017n01} formally
corresponds to taking into account the Feynman graphs depicted in
\Fig{fig-coul-1}, where graphs (a) and (b) represent 
the first and  second terms in the right side of \Eq{eqn181017n05}.

%
%
%
%
\begin{figure}[h]
\includegraphics[width=26pc]{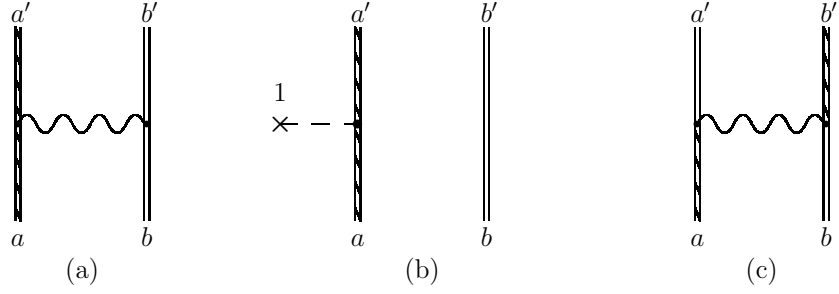}
\caption{
The direct (a) and exchange (c) Feynman graphs representing the one-photon exchange.
The double line describes electron in the Furry picture with potential
$V^\txt{F}= -\alpha Z/r$,
the double striped line describes electron in the Furry picture with potential
$V^\txt{F}_\txt{cont}= -\alpha (Z-1)/r$.
The graph (b) represents interaction of the incident electron with the potential
$\Delta V_\txt{cont}= -\alpha/r$.
}
\label{fig-coul-2}
\end{figure}
%
%

For the derivation of the amplitude for the resonant scattering channel 
(see \Br{eqn170705n02}) 
the Feynman graphs depicted in \Fig{fig-coul-2} 
have to be taken into account.
The graphs (a) and (c) present the direct and exchange graphs, respectively, of the one-photon exchange.
The double line describes an electron in the Furry picture with the potential $V^\txt{F}= -\alpha Z/r$,
the double striped line describes continuum  electrons in the Furry picture with the potential 
$V^\txt{F}_\txt{cont}= -\alpha (Z-1)/r$.
The graph (b) represents the interaction of the incident electron 
with the potential
$\Delta V_\txt{cont}=V^\txt{F}-V^\txt{F}_\txt{cont}= -\alpha/r$ 
(i.e. with the potential of the `remaining' charge of the nucleus).

The contribution of the graph (a) (or the graph (c)) from 
\Fig{fig-coul-2}
is given by \cite{andreev08pr}
\begin{eqnarray}
(I)_{u_1 u_2 d_1 d_2}
&=&\label{eqn1906071624}
\alpha
\int d\zhr_1 d\zhr_2\,
\bpsi_{u_1}(\zhr_1)
\bpsi_{u_2}(\zhr_2)
\gamma^{\mu_1}_1 \gamma^{\mu_2}_2
I_{\mu_1 \mu_2}(|\Omega|, r_{12})
\psi_{d_1}(\zhr_1)
\psi_{d_2}(\zhr_2)
\,,
\end{eqnarray}
where $r_{12}=|\zhr_1-\zhr_2|$ and
\begin{eqnarray}
I^\txt{c}_{\mu_1 \mu_2}
&=&
\frac{\delta_{\mu_1 0} \delta_{\mu_2 0}}{r_{12}}
\,,
\\
I^\txt{t}_{\mu_1 \mu_2}
&=&
-\left(
\frac{\delta_{\mu_1 \mu_2}}{r_{12}} e^{i\Omega r_{12}}
+
\frac{\partial}{\partial x^{\mu_1}_1}
\frac{\partial}{\partial x^{\mu_2}_2}
\frac{1-e^{i\Omega r_{12}}}{r_{12}\Omega^2}
\right)
(1-\delta_{\mu_1 0})(1-\delta_{\mu_2 0})
\,.
\end{eqnarray}

While the exchange contribution (graph (c)) does not contain any divergence, the direct contribution (graph (a)) does. 
However, the contribution to the amplitude 
given by graph (b) is also divergent and it turns out 
that the divergences in graphs (a) and (b) cancel each other.

Let us consider this point more in detail. 
First of all we note that the problem with graph (a) arises only 
due to the Coulomb part of the transition matrix element. 
We now assume that $d_1=u_1=e_{\zhp,\mu}$
[the incident and scattered electron having the same energy 
($\varepsilon$)] and $d_2=u_2=1s$ (the $1s$-electron) 
and consider only the above mentioned 
part $I^\txt{c}$ of the transition matrix element 
\begin{eqnarray}
(I^\txt{c})_{u_1 u_2 d_1 d_2}
&=&
\alpha
\int d\zhr_1 d\zhr_2
\psi_{u_1}^{+}(\zhr_1)
\psi_{u_2}^{+}(\zhr_2)
\frac{1}{r_{12}}
\psi_{d_1}(\zhr_1)
\psi_{d_2}(\zhr_2)
\,.
\end{eqnarray}
Using the expansion 
\cite{varshalovich}
\begin{eqnarray}
\frac{1}{r_{12}}
&=&\label{eqnr12}
\sum\limits_{K=0}^{\infty}  \frac{r_{<}^K}{r_{>}^{K+1}} P_{K}(\cos\theta)
\,,
\end{eqnarray}
where $r_{<}=\min(r_1,r_2)$, $r_{>}=\max(r_1,r_2)$
and retaining only the term with $K=0$ (the terms with $K>0$ have no divergences) we can write 
\begin{eqnarray}
(I^\txt{c}_0)_{u_1 u_2 d_1 d_2}
&=\label{eqn181030n01}
\alpha
\int d\zhr_1 d\zhr_2
\psi_{u_1}^{+}(\zhr_1)
\psi_{u_2}^{+}(\zhr_2)
\frac{1}{r_{>}}
\psi_{d_1}(\zhr_1)
\psi_{d_2}(\zhr_2)
\,.
\end{eqnarray}
The wave function of an electron with a given momentum 
and polarization can be decomposed into the complete set 
of partial waves with a certain energy and angular momentum  (see 
\Eq{eqn181017n02}). The divergence appears only for  
the case of identical partial waves for the incident and scattered electrons, which we shall now consider.
The wave function of $1s$-electron decreases exponentially, 
accordingly, the integration over $\zhr_2$ is convergent.
Asymptotically (when $r\to\infty$) the upper and lower components 
of the incident electron wave function are given by 
\begin{eqnarray}
g_{\varepsilon jlm}(r)
&\sim&
C_g \cos(pr+\xi\log{2pr}+\delta_{jl})
\,,
\\
f_{\varepsilon jlm}(r)
&\sim&
C_f \sin(pr+\xi\log{2pr}+\delta_{jl})
\,,
\end{eqnarray}
where $\xi=\alpha Z\varepsilon/p$ is the Sommerfeld parameter, $\delta_{jl}$ are the phase shifts,
$C_g$ and $C_f$ are constants
\cite{akhiezer65b}.
Accordingly, the integration over $\zhr_1$ contains a divergent part  at the upper limit
($r_1\to\infty$)
\begin{eqnarray}
\int dr_1
\frac{1}{r_1}
\left[
|C_g|^2\cos^2(pr_1+\xi\log{2pr_1}+\delta_{jl})
+|C_f|^2\sin^2(pr_1+\xi\log{2pr_1}+\delta_{jl})
\right]
\,.
\end{eqnarray}
After a regularization of the integral over $r_1$ the divergent part of
\Eq{eqn181030n01}
can be singled out and, as will be shown, it will be exactly cancelled by 
the divergence contained in the contribution to the amplitude due to 
the interaction of the continuum electron with the potential 
$\Delta V_\txt{cont}$.

It is convenient to split the integral $I^\txt{c}_0$ into two parts
\begin{eqnarray}
(I^\txt{c}_0)_{u_1 u_2 d_1 d_2}
&=&\nonumber
\alpha
\int d\zhr_1 d\zhr_2
\left(\frac{1}{r_{>}}
-\frac{1}{r_{1}}\right)
\langle\psi_{u_1}^{+}(\zhr_1)
\psi_{d_1}(\zhr_1)\rangle
\langle\psi_{u_2}^{+}(\zhr_2)
\psi_{d_2}(\zhr_2)\rangle
\\
&&\nonumber
+
\alpha
\int d\zhr_1 d\zhr_2
\left(\frac{1}{r_{1}}\right)
\langle\psi_{u_1}^{+}(\zhr_1)
\psi_{d_1}(\zhr_1)\rangle
\langle\psi_{u_2}^{+}(\zhr_2)
\psi_{d_2}(\zhr_2)\rangle
\\
&=&\nonumber
\alpha
\int d\zhr_2 \int\limits_{|\zhr_1|<|\zhr_2|} d\zhr_1
\left(\frac{1}{r_{2}}
-\frac{1}{r_{1}}\right)
\langle\psi_{u_1}^{+}(\zhr_1)
\psi_{d_1}(\zhr_1)\rangle
\langle\psi_{u_2}^{+}(\zhr_2)
\psi_{d_2}(\zhr_2)\rangle
\\
&&
+
\alpha
\int d\zhr_1
\left(
\frac{1}{r_{1}}\right)
\langle\psi_{u_1}^{+}(\zhr_1)
\psi_{d_1}(\zhr_1)\rangle
\,
\delta_{u_2 d_2}
\,. 
\end{eqnarray}
Here we used the fact that the wave function 
of the $1s$-electron is normalized to unity.
The integral containing the term 
$ (1/r_{2} -1/r_{1} )$ is obviously convergent.   
The integral with $ 1/r_{1} $ is divergent but is 
exactly cancelled out by the divergent part of 
the contribution to the amplitude due to 
the interaction $\Delta V_\txt{cont}$ of the incident electron 
with the `remaining' charge, which reads
\begin{eqnarray}
(\Delta V_\txt{cont})_{u_1 u_2 d_1 d_2}
&=&\label{eq567jdfnv}
\alpha
\int d\zhr_1
\psi_{u_1}^{+}(\zhr_1)
\left(\frac{-1}{r_{1}}\right)
\psi_{d_1}(\zhr_1)
\,
\delta_{u_2 d_2}
\,.
\end{eqnarray}

The consideration given in the previous paragraphs is based 
on the decomposition $ Z = (Z - 1) + 1$. Its first term 
corresponds to the effective charge of the nucleus 
(as seen asymptotically by the incident and scattered electron) 
screened by the bound electron which essentially 
is regarded as a point-like charge $(-1)$ placed at the origin. 
Using the Furry picture the Coulomb interaction between 
the continuum electron and such a point-like electron 
is taken into account to all orders. 
At the same time the difference between this interaction and the Coulomb interaction between the continuum electron and the bound electron in the $1s$-state is considered in the first order. 
We note that this difference is of short range 
corresponding to the interaction 
with an electrically neutral system, which is considered using the perturbation theory. Indeed, the sum of $I^\txt{c}_0$ 
and $\Delta V_\txt{cont}$
\begin{eqnarray}
(I^\txt{c}_0+\Delta V_\txt{cont})_{u_1 u_2 d_1 d_2}
&=&\label{coulomb-screening}
\alpha
\int d\zhr_2 \int\limits_{|\zhr_1|<|\zhr_2|} d\zhr_1
\left(\frac{1}{r_{>}}
-\frac{1}{r_{1}}\right)
\langle\psi_{u_1}^{+}(\zhr_1)
\psi_{d_1}(\zhr_1)\rangle
\langle\psi_{u_2}^{+}(\zhr_2)
\psi_{d_2}(\zhr_2)\rangle
\end{eqnarray}
is finite and represents a correction due to interaction with a short-range potential. Within our approach
the graph (a) in \Fig{fig-coul-2} 
must be always considered together with graph (b). 
 
For light ions the accuracy of our approach may be improved if we replace  the interaction with a point-like charge by the interaction with the charge density given by the $1s$-electron wave function and adjust  
the Furry picture accordingly. The influence of this replacement on the results is discussed in the next section (Figs. \ref{figure07} and \ref{figure07_2}).

\vspace{0.15cm} 
The Feynman graph depicted in
\Fig{fig-coul-2} (c)
represents the exchange graph of the one-photon exchange,
it is finite and does not require a special treatment. 
\vspace{0.15cm} 

Up to now we considered the direct (non-resonant) elastic scattering. 
Let us now briefly discuss the description of the resonant channel of the scattering process which becomes relevant when 
the sum of the energies of the incident and the bound electrons 
is close to the energy of a doubly excited (autoionizing) state. 
In such a case the contribution of two- and more-photon exchange 
between electrons in low-lying states becomes of importance. 
Moreover, the doubly excited states 
are normally quasidegenerate, hence, 
a perturbation theory for quasidegenerate states 
has to be used. For this purpose the LPA \cite{andreev08pr} 
is employed.

For application of the quasidegenerate perturbation theory within the LPA
we introduce the set of two-electron configurations ($g$) in the $j-j$ coupling scheme
which includes all two-electron configurations composed by
a certain set of electrons (for example, $1s,2s,2p,3s,3p,3d$ electrons)
\begin{eqnarray}
\Psi^{(0)}_{JMj_1j_2l_1l_2n_1n_2}(\zhr_1,\zhr_2)
&=&\label{jjcs}
N'\sum\limits_{m_1m_2} C^{j_1j_2}_{JM}(m_1,m_2)
\det\{\psi_{n_1j_1l_1m_1}(\zhr_1),\psi_{n_2j_2l_2m_2}(\zhr_2)\}
\,,
\end{eqnarray}
where $j$ and $m$ is the total angular momentum and its projection,
$l$ is the orbital angular momentum defining the parity,
$n$ denotes the principle quantum number or energy (for the continuum electrons),
$J$ and $M$ are the total angular momentum of the two-electron configuration
and its projection,
$N'$ is the normalizing constant.

In the LPA a matrix $V$ 
is introduced, which is determined by the one and two-photon exchange,
electron self-energy and vacuum polarization matrix elements and other QED corrections \cite{andreev08pr} and which can be derived order by order within the QED perturbation theory.
The matrix $V = V^{(0)} + \Delta V$ is considered as a block matrix
\begin{eqnarray}
V
&=&
        \left[
        \begin{array}{cc}
        V_{11} & V_{12}  \\
        V_{21}  &V_{22}  \\
        \end{array}
        \right]
\,=\,
        \left[
        \begin{array}{cc}
       V^{(0)}_{11}+ \Delta V_{11} & \Delta V_{12}  \\
        \Delta V_{21}  &V^{(0)}_{22}+\Delta V_{22}  \\
        \end{array}
        \right]
\,.
\end{eqnarray}
The matrix $V_{11}$ is defined on the set $\setg$,
which contains configurations mixing with the reference state
$n_g$ $\in$ $g$.
The matrix $V^{(0)}$ is a diagonal matrix composed of the sum of the one-electron Dirac energies. 
The matrix $\Delta V_{11}$ is not a diagonal matrix, but it contains a  small parameter of the QED perturbation theory.
The matrix $V_{11}$ is a finite-dimensional matrix 
and can be diagonalized  numerically:
\begin{eqnarray}
V^\txt{diag}_{11}
&=&\label{matrixv11}
B^{t}
{V_{11}}
B
\,,
\phantom{12345}
B^{t}B=I
\,.
\end{eqnarray}
The matrix $B$ defines a  transformation 
to the basis set in which
the matrix $V_{11}$ is a diagonal matrix.
Then, the standard perturbation theory can be applied for the diagonalization of
the infinite-dimensional matrix $V$.
The eigenvectors of $V$ read
\cite{andreev08pr}
\begin{eqnarray}
\Phi_{ n_g}
&=&\label{eigenphi}
\sum\limits_{ k_g \in g}
B_{k_g {n_g}}\Psi^{(0)}_{ k_g}
+
\sum\limits_{ k\notin g, { l_g\in g}}
[\Delta V_{21}]_{ k { l_g}}
\frac{B_{{l_g} { n_g}}}
{E^{(0)}_{n_g}-E^{(0)}_{k}}
\Psi^{(0)}_{k}
+
\cdots
\,,
\end{eqnarray}
where $n_g\equiv(JMj_1j_2l_1l_2n_1n_2)$ is a complete set of quantum numbers describing 
the reference state,
indices $k$, $l_g$ denote the two-electron configurations:
the index $l_g$ runs over all configurations of the set $g$;
the index $k$ runs over all configurations not included in the set $g$
(this implies the integration over the positive- and negative energy continuum).
Here, $E^{(0)}_k$ is the energy of the two-electron configuration $\Psi^{(0)}_k$ given by the sum of the one-electron Dirac energies.

\par
The amplitude of the scattering process is given as a matrix element 
of the operator ($\hXi$)
\begin{eqnarray}
U^\txt{Auger}_{if}
&=&\label{ampl}
\langle\Psi^{(0)}_\txt{fin}|\hXi|\Phi_\txt{ini}\rangle
\,.
\end{eqnarray}
The bra vector corresponds to the wave function 
describing noninteracting electrons \Eq{jjcs},
the ket vector is given by the eigenvector 
\Eq{eigenphi}.
The operator $\hXi$ is derived within the QED perturbation theory order by order
\cite{andreev08pr,andreev09p042514}. 
In the first and second orders of the perturbation theory 
it is represented by Feynman graphs depicted in
\Fig{fig-opx}
and
\ref{fig-tpx},
respectively.

The total amplitude of the process
\Br{eqn170705n01}, including its resonant part \Br{eqn170705n01}, is given by 
\begin{eqnarray}
U_{if}
&=&
U^\txt{Coul}_{if}
+
U^\txt{Auger}_{if}
\,, 
\end{eqnarray}
where the Coulomb and Auger contributions are 
given by \Eq{eqn181017n01} and \Br{ampl}, respectively. 

The numerical calculation of the Coulomb amplitude  $U^\txt{Coul}_{if}$ is discussed in
Appendix~\ref{appendixb}.

The transition probability is expressed via the amplitude according to 
\cite{akhiezer65b}
\begin{eqnarray}
dw_{if}
&=&\label{eqn181026n09}
2\pi |U_{if}|^2 \delta(\mee_i-\mee_f)
\frac{d^3\zhp_f}{(2\pi)^3}
\,,
\end{eqnarray}
where $\mee_i$, $\mee_f$ are the energies of the initial and final states of the system and $\zhp_f$ is the momentum of the scattered electron.

The cross section is defined as
\begin{eqnarray}
d\sigma_{if}
&=&\label{eqn181026n10}
\frac{dw_{if}}{j}
\,,
\end{eqnarray}
where $j=p_i/\varepsilon_i$ is the flux of the incident electrons 
having an energy $\varepsilon_i$ and a momentum $p_i$.
Accordingly, the double and single differential cross sections 
for elastic electron scattering read 
\begin{eqnarray}
\frac{d\sigma_{if}}{d\varepsilon_f d\Omega_f} (\varepsilon_f, \theta_f)
&=&
2\pi |U_{if}|^2 \delta(\varepsilon_f - \varepsilon_i)\frac{\varepsilon_i}{p_i}\frac{p_f \varepsilon_f}{(2\pi)^3}
\,,
\end{eqnarray}

\begin{eqnarray}
\frac{d\sigma_{if}}{ d\Omega_f}(\varepsilon_f=\varepsilon_i, \theta_f)
&=&\label{eq567hs}
2\pi |U_{if}|^2 \frac{\varepsilon_i}{p_i}\frac{p_f \varepsilon_f}{(2\pi)^3}
\,,
\end{eqnarray}
where $\varepsilon_f$ and $\Omega_f$ are the energy and solid angle 
(with polar angle $\theta_f$) of the scattered electron, respectively.

\section{Results and discussion}
In this section we discuss results of applications of our theory to electron scattering on hydrogen-like ions ranging from boron to uranium. The calculated cross sections will be given in the rest frame of the ion and for those impact energies where only $LL$ autoionizing states participate in the scattering process. Since the Coulomb scattering is especially strong in forward angles, we restrict ourselves to the consideration of electron scattering to backward angles for which the Coulomb contribution is minimal. 

In experiments on electron-ion scattering free electrons are often replaced by electrons bound in light atomic (molecular) targets which serve as a source of (quasi-)free electrons. Therefore, in what follows we consider collisions of hydrogen-like ions not only with free electrons but also with molecular hydrogen.

%
%
\begin{figure}[h]
\includegraphics[width=35pc]{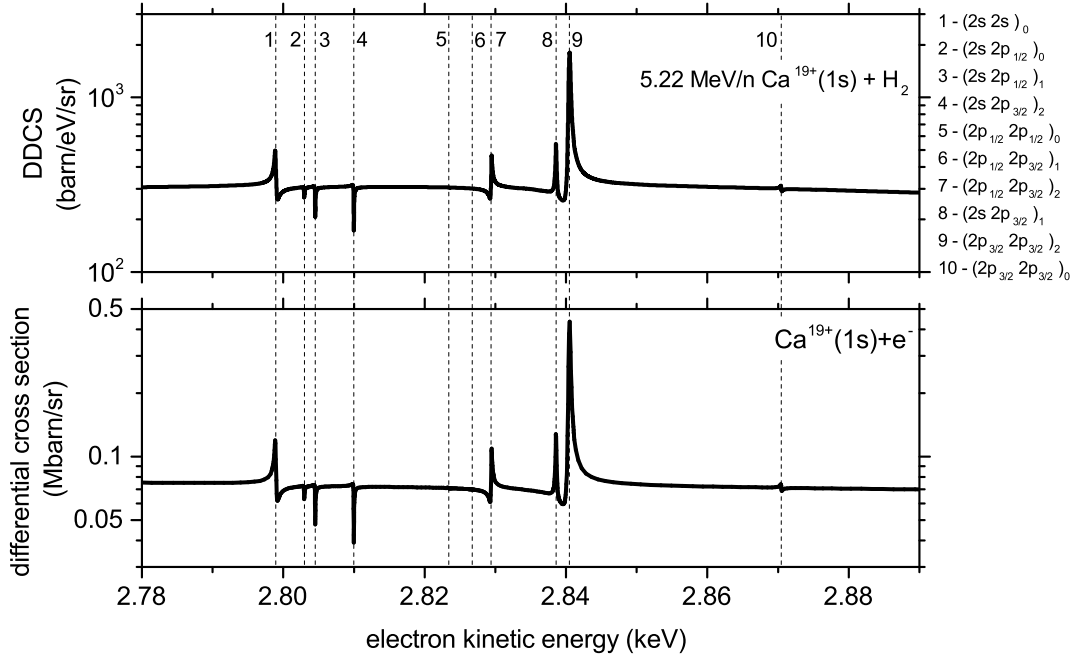}
\caption{Double differential cross section for the elastic electron scattering in collisions of  5.22 MeV/n Ca$^{19+}(1s)$ with H$_2$ target (top) and single differential cross section for the elastic electron scattering in collisions of Ca$^{19+}(1s)$ with free electrons (bottom).
}
\label{figure08}
\end{figure}
%
%
\begin{figure}[h]
\includegraphics[width=35pc]{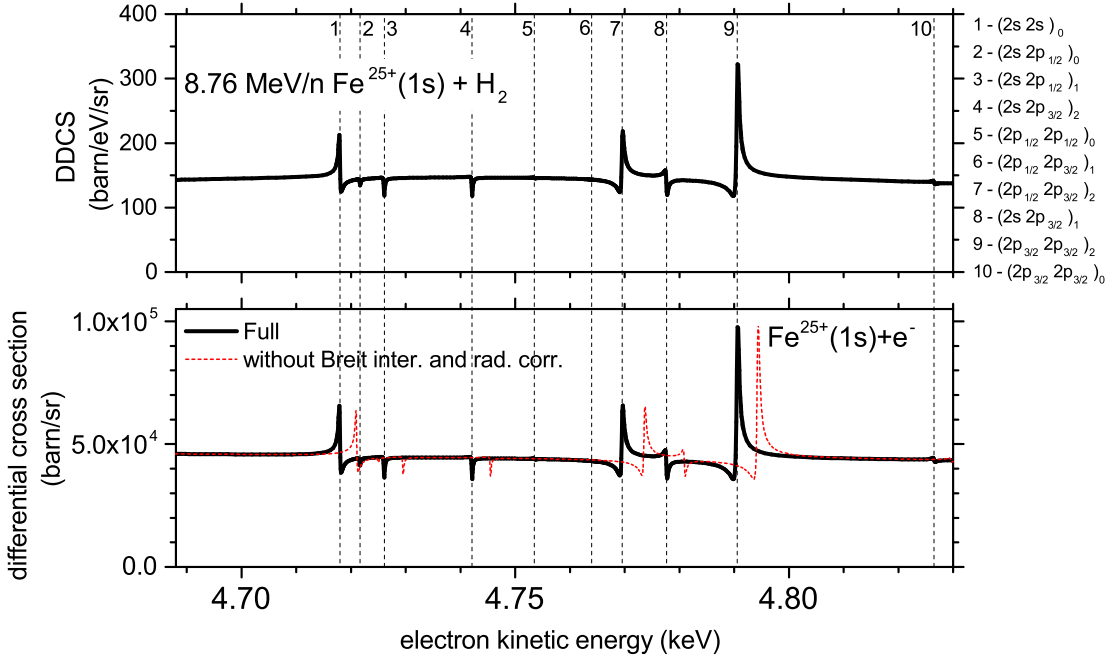}
\caption{The same as in Fig.~\ref{figure08} but for 8.76 MeV/n Fe$^{25+}(1s)$.
The solid black lines represent the results of the exact QED calculation,
the dashed red line corresponds to the calculation with disregard of the Breit
part of the interelectron interaction and the real part of the radiative corrections.
}
\label{figure09}
\end{figure}
%
%
\begin{figure}[h]
\includegraphics[width=35pc]{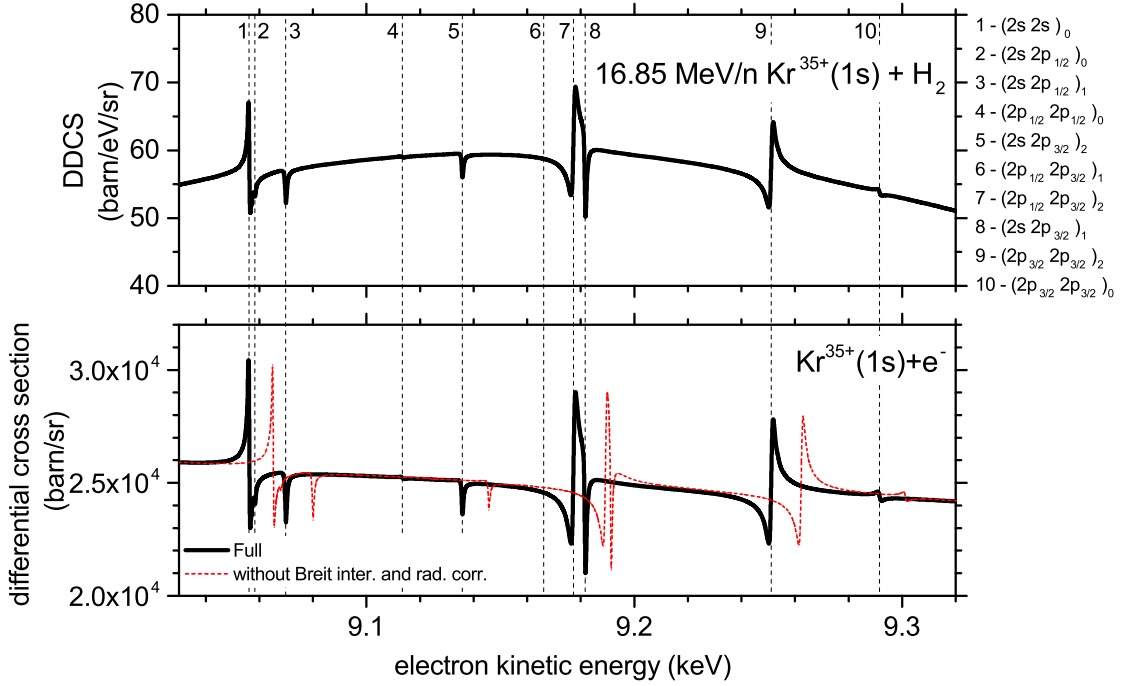}
\caption{The same as in Fig.~\ref{figure09} but for 16.85 MeV/n Kr$^{35+}(1s)$.
}
\label{figure10}
\end{figure}
%
%
\begin{figure}[h]
\includegraphics[width=30pc]{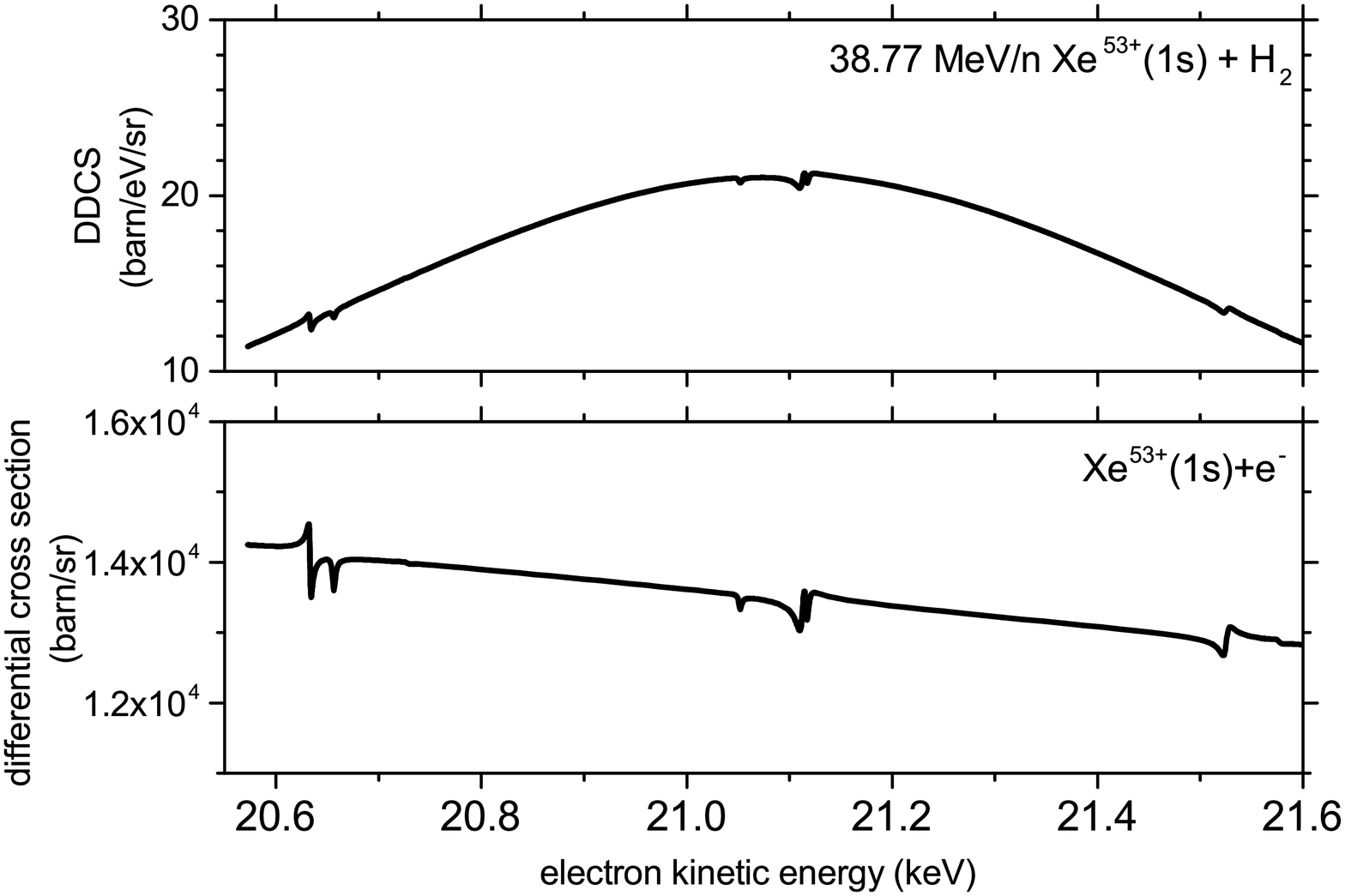}
\caption{The same as in Fig.~\ref{figure08} but for 38.77 MeV/n Xe$^{53+}(1s)$.
}
\label{figure11}
\end{figure}
%
%

In the bottom panels of Figs.~\ref{figure08}~--~\ref{figure11} the single differential cross section for scattering of free electrons is presented as a function of the scattered electron kinetic energy for Ca$^{19+}(1s)$, Fe$^{25+}(1s)$, Kr$^{35+}(1s)$ and Xe$^{53+}(1s)$. In these figures we observe a smooth background, caused by Coulomb scattering, which is superimposed by maxima and minima arising due to the resonant scattering as well as interference between the resonant and the Coulomb parts of the scattering process.
The figures show that the differential cross section strongly depends on the charge of the ionic nucleus ($Z$) both qualitatively and quantitatively.

The total scattering amplitude can be written as a sum of the two parts:  the non-resonant (Coulomb) part and the resonant part. Accordingly the  cross section can be split into three terms corresponding to the Coulomb scattering, the resonant scattering and their interference.

\begin{figure}[h]
\includegraphics[width=35pc]{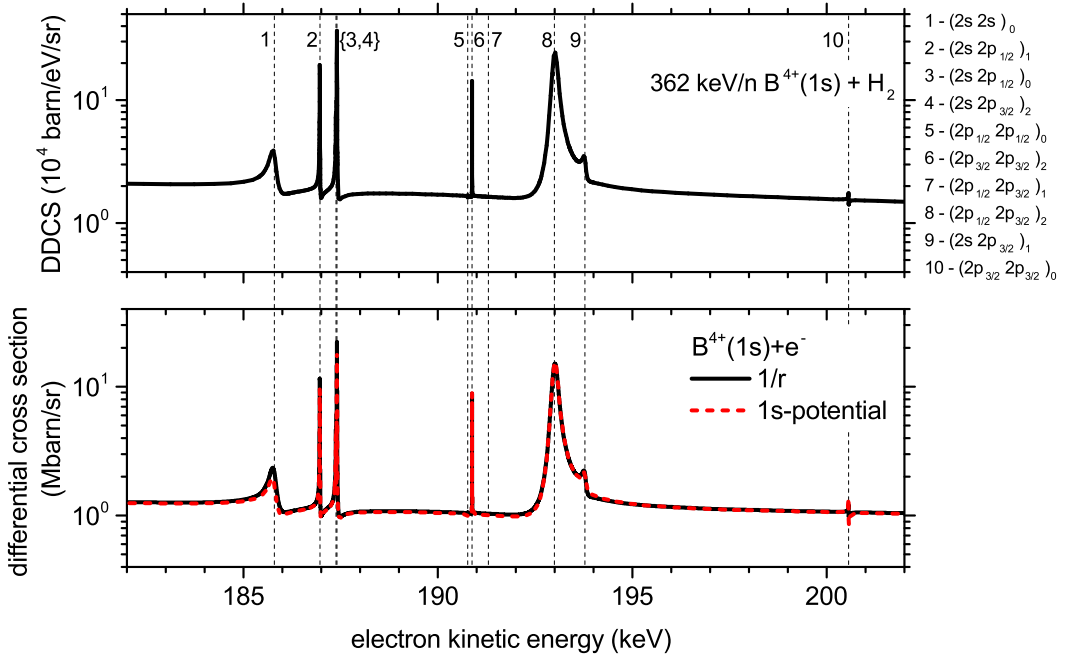}
\caption{
Same as in Fig.~\ref{figure08} but for 0.362 MeV/n B$^{4+}(1s)$.
The solid black curves represent results of calculations with potential
$\Delta V_\txt{cont}=-\alpha/r$,
the dotted red curve corresponds to the calculation with the potential $\Delta V_\txt{cont}$
given by
\Eq{eqn190507n01}.
}
\label{figure07}
\end{figure}

When the charge $Z$ of the ionic nucleus varies, the amplitude for Coulomb scattering in the vicinity of resonances effectively scales as $1/Z$. Further, the Schr\"odinger equation predicts that, provided the total width of an autoionizing state is determined mainly by the electron-electron interaction, the amplitude for resonant scattering scales as $1/Z$ as well. Therefore, for not too heavy ions one could expect that all the contributions to the cross section -- the Coulomb and the resonant parts and the interference term -- scale with $Z$ roughly as $1/Z^2$. Indeed, our calculations for ions ranging between $Z\sim5$ and $Z\sim20$ are in qualitative agreement with this scaling (for illustration see Figs.~\ref{figure08} and \ref{figure07}). For heavier ions the amplitude for the resonant scattering begins to decrease with $Z$ faster than $1/Z$. This is mainly caused by a rapid growth of the radiative contribution to the total widths which in heavy ions outperforms the Auger decay.
Accordingly, with increasing $Z$ the contribution of the resonant scattering decreases faster than the contribution from the interference term. Whereas for relatively small $Z$ all the three parts of the cross section are equally important, for higher $Z$ the purely resonant part becomes of minor importance and the resonance structure in the cross section is determined solely by the interference term. 

We also note that the order of resonances also depends on $Z$, in particular, it leads to different orders of maxima-minima for different $Z$. The resonant energies of the incident electron as well as the energies and widths of the corresponding autoionizing states for various ions are presented in Tables~\ref{tab1}~--~\ref{tab4}. 
%
%
\begin{table}
 \caption{\label{tab1}The parameters characterizing the resonant structure
of the cross section for various autoionizing ($LL$) states of helium-like calcium are presented.
The second and third columns present the energies ($E^{'}=E_{(LL)}-2mc^2$) and widths ($\Gamma$)  of the autoionizing states.
The forth column present the resonant energies of the scattered electron ($\varepsilon_\txt{res}=E_{(LL)}-\varepsilon_{1s}$)
The data are given in the rest frame of the ion.}
\begin{tabular}{cccc}
\hline
autoionizing & $E^{'}$& $\Gamma$ & $\varepsilon_\txt{res}$ \\
state & keV& eV& keV\\
\hline
$(  2s_{    })^{2}_{ 0}$&       -2.6710&          0.25&        2.7989\\
$(  2s_{    }  2p_{1/2 })_{ 0}$&       -2.6670&          0.08&        2.8030\\
$(  2s_{    }  2p_{1/2 })_{ 1}$&       -2.6654&          0.08&        2.8045\\
$(  2s_{    }  2p_{3/2 })_{ 2}$&       -2.6600&          0.07&        2.8100\\
$(  2p_{1/2 })^{2}_{ 0}$&       -2.6465&          0.13&        2.8234\\
$(  2p_{1/2 }  2p_{3/2 })_{ 1}$&       -2.6432&          0.13&        2.8267\\
$(  2p_{1/2 }  2p_{3/2 })_{ 2}$&       -2.6405&          0.16&        2.8294\\
$(  2s_{    }  2p_{3/2 })_{ 1}$&       -2.6313&          0.19&        2.8386\\
$(  2p_{3/2 })^{2}_{ 2}$&       -2.6295&          0.33&        2.8405\\
$(  2p_{3/2 })^{2}_{ 0}$&       -2.5995&          0.12&        2.8705\\
\hline
\end{tabular}
\end{table}
%
%
\begin{table}
 \caption{\label{tab2} Same as in Table \ref{tab1} but for helium-like iron. }
\begin{tabular}{cccc}
\hline
autoionizing & $E^{'}$& $\Gamma$ & $\varepsilon_\txt{res}$ \\
state & keV& eV& keV\\
\hline
$(  2s_{    })^{2}_{ 0}$&       -4.5597&          0.31&        4.7179\\
$(  2s_{    }  2p_{1/2 })_{ 0}$&       -4.5560&          0.21&        4.7216\\
$(  2s_{    }  2p_{1/2 })_{ 1}$&       -4.5516&          0.20&        4.7260\\
$(  2s_{    }  2p_{3/2 })_{ 2}$&       -4.5356&          0.19&        4.7421\\
$(  2p_{1/2 })^{2}_{ 0}$&       -4.5242&          0.35&        4.7534\\
$(  2p_{1/2 }  2p_{3/2 })_{ 1}$&       -4.5138&          0.37&        4.7639\\
$(  2p_{1/2 }  2p_{3/2 })_{ 2}$&       -4.5082&          0.45&        4.7695\\
$(  2s_{    }  2p_{3/2 })_{ 1}$&       -4.5001&          0.31&        4.7776\\
$(  2p_{3/2 })^{2}_{ 2}$&       -4.4872&          0.53&        4.7905\\
$(  2p_{3/2 })^{2}_{ 0}$&       -4.4513&          0.34&        4.8264\\
\hline
\end{tabular}
\end{table}
%
%
\begin{table}
 \caption{\label{tab3} Same as in Table \ref{tab1} but for helium-like krypton.}
\begin{tabular}{cccc}
\hline
autoionizing & $E^{'}$& $\Gamma$ & $\varepsilon_\txt{res}$ \\
state & keV& eV& keV\\
\hline
$(  2s_{    })^{2}_{ 0}$&       -8.8808&          0.56&        9.0553\\
$(  2s_{    }  2p_{1/2 })_{ 0}$&       -8.8783&          0.73&        9.0579\\
$(  2s_{    }  2p_{1/2 })_{ 1}$&       -8.8672&          0.73&        9.0689\\
$(  2p_{1/2 })^{2}_{ 0}$&       -8.8238&          1.15&        9.1124\\
$(  2s_{    }  2p_{3/2 })_{ 2}$&       -8.8008&          0.69&        9.1353\\
$(  2p_{1/2 }  2p_{3/2 })_{ 1}$&       -8.7704&          1.38&        9.1657\\
$(  2p_{1/2 }  2p_{3/2 })_{ 2}$&       -8.7593&          1.51&        9.1768\\
$(  2s_{    }  2p_{3/2 })_{ 1}$&       -8.7553&          0.80&        9.1808\\
$(  2p_{3/2 })^{2}_{ 2}$&       -8.6855&          1.47&        9.2506\\
$(  2p_{3/2 })^{2}_{ 0}$&       -8.6456&          1.34&        9.2905\\
\hline
\end{tabular}
\end{table}
%
%
\begin{table}
 \caption{\label{tab4} Same as in Table \ref{tab1} but for helium-like xenon.}
\begin{tabular}{cccc}
\hline
autoionizing & $E^{'}$& $\Gamma$ & $\varepsilon_\txt{res}$ \\
state & keV& eV& keV\\
\hline
$(  2s_{    }  2p_{1/2 })_{ 0}$&      -20.669&          3.48&       20.632\\
$(  2s_{    })^{2}_{ 0}$&      -20.669&          2.18&       20.632\\
$(  2s_{    }  2p_{1/2 })_{ 1}$&      -20.646&          3.46&       20.655\\
$(  2p_{1/2 })^{2}_{ 0}$&      -20.574&          4.94&       20.727\\
$(  2s_{    }  2p_{3/2 })_{ 2}$&      -20.250&          3.41&       21.051\\
$(  2p_{1/2 }  2p_{3/2 })_{ 1}$&      -20.205&          6.79&       21.096\\
$(  2p_{1/2 }  2p_{3/2 })_{ 2}$&      -20.189&          6.94&       21.112\\
$(  2s_{    }  2p_{3/2 })_{ 1}$&      -20.185&          3.51&       21.116\\
$(  2p_{3/2 })^{2}_{ 2}$&      -19.777&          6.88&       21.524\\
$(  2p_{3/2 })^{2}_{ 0}$&      -19.725&          6.80&       21.575\\
\hline
\end{tabular}
\end{table}
%
%

In order to investigate the influence of the Breit interaction and the radiative corrections (self-energy and vacuum polarization) we performed calculations where these corrections were omitted.
The red dotted lines in the bottom panels of Figs.~\ref{figure09} and \ref{figure10} correspond to calculations in which the Breit interaction and the radiative corrections were neglected. The energy shift of the resonances is clearly visible and roughly equals to $\sim 4$ eV and $\sim 10$ eV for iron and krypton, respectively. A slight decrease in the differential cross section due to the contribution of the Breit interaction is noticeable but turns out to be rather small even for krypton.
 
In the top panels of Fig.~\ref{figure08}~--~\ref{figure11} we present  results for collisions of highly charged ions with $H_2$. In order to evaluate the doubly differential scattering cross section for collisions with molecular hydrogen we use the Impulse Approximation where the electrons, which are initially bound in hydrogen, are considered as quasi-free. Using this approximation one can show that 
the cross section in collisions with hydrogen can be expressed via 
the cross section in collisions with free electrons according to 
\begin{eqnarray}
\left(\frac{d^2\sigma (\varepsilon_f,\theta_f)}{d\varepsilon_f d\Omega_f}\right)_{H_2}
&=&\label{eqn789jd}
2\left(\frac{d\sigma (\varepsilon_f,\theta_f)}{ d\Omega_f} \right)_{\txt{free}} \frac{\varepsilon_f}{\gamma p_f} \, J\left(p'_z\right)
\,. 
\end{eqnarray}
Here, $\gamma=1/{\sqrt{1-V_{\txt{col}}^2}}$ and 
$p'_z = \gamma(p_f-V_{\txt{col}}\varepsilon_f)$, where  
$V_{\txt{col}}$ is the collision velocity. 
Further, 
\begin{eqnarray}
J(p'_z)
&=&\nonumber
\int d^2{\bf{p}}_{\perp}' |\psi_{1s}({\bf{p}}_{\perp}',p'_z)|^2\\
&=&
\frac{8}{3 \, \pi}\frac{\alpha^5}{(\alpha^2+p'^2_z)^3}
\,,\label{eqn789jd34}
\end{eqnarray}
is the Compton profile for the $1s$-electron of hydrogen atom with the corresponding wave function $\psi_{1s}({\bf{p}}')$, 
${\bf{p}}'=({\bf{p}}_{\perp}',p'_z)$ is the momentum of the incident electron in the rest frame of hydrogen, $\alpha$ is the hyper-fine structure constant (which corresponds to the characteristic orbiting momentum of the $1s$-electron in hydrogen expressed in relativistic units) and $\left(\frac{d\sigma (\varepsilon_f,\theta_f)}{ d\Omega_f} \right)_{\txt{free}}$ is the cross section for collision with a free electron given by Eq.~(\ref{eq567hs}).

Comparing the cross sections for collisions with free and quasi-free electrons we can conclude that the resonance structure remains basically the same and the only difference is the bending of the background caused by the convolution with the Compton profile. The bending is more prominent for collisions with heavier ions since the energy interval considered scales with $Z$ as $Z^2$. 

In our approach the Coulomb interaction of the incident and scattered electron with the nucleus and partly with the bound $1s$-electron 
is taken into account within the Furry picture.
In \Eq{eq567jdfnv} the interaction of the electron in the continuum 
with the $1s$-electron is considered -- in the first order of the  perturbation theory -- as the Coulomb interaction with 
a point-like charge located in the origin. 
This is consistent with the Furry picture for the continuum electrons, 
in which they are regarded as moving in the field $V^\txt{F}_\txt{cont}=-\alpha(Z-1)/r$. \Eq{coulomb-screening}
represents the difference between the long-range term of
the Coulomb interaction of the continuum electron 
with the $1s$-electron, given by \Eq{eqn181030n01} 
(contribution of the term with $K=0$ in
\Eq{eqnr12}), 
and the interaction of the continuum electron with 
the point-like charge placed in the origin. Hence,
\Eq{coulomb-screening}
describes the interaction with a short-range potential
in the first order of the perturbation theory.
We note that all numerical results presented in
Figs.~\ref{figure08}-\ref{figure11}
were obtained using the Furry picture mentioned above.  

In order to investigate the importance of the higher orders of the perturbation theory corresponding to the interaction with this potential,
we replaced in Eqs.~\Br{eq567jdfnv} and \Br{coulomb-screening} the interaction with a point-like charge by the interaction with a charge density corresponding to the $1s$-electron wave function
\begin{eqnarray}
\Delta V_\txt{cont}
&=&\label{eqn190507n01}
-\alpha
\int d\zhr'
\frac{\psi^{+}_{1s}(\zhr')\psi_{1s}(\zhr')}{|\zhr'-\zhr|}
\,. 
\end{eqnarray} 
Since the $1s$-electron wave-function is independent of angular variables,
with employment of the decomposition
\Eq{eqnr12}
the potential
\Eq{eqn190507n01}
can be deduced to
\begin{eqnarray}
\Delta V_\txt{cont}(r)
&=&\label{eqn190507n01-1906071558}
-\alpha
\int^{\infty}_0 dr'\,
\frac{\rho(r')}{r_{>}}
\,,
\end{eqnarray} 
where $\rho(r')$ is the probability density function of the $1s$-electron, $r_{>}=\max(r,r')$.

With this replacement the Furry picture should be changed accordingly:
now the continuum electron is considered to be moving in the potential
\begin{eqnarray}
V^\txt{F}_\txt{cont}
&=&\label{eqn190507n02}
-\frac{\alpha Z}{r}
+\alpha
\int d\zhr'
\frac{\psi^{+}_{1s}(\zhr')\psi_{1s}(\zhr')}{|\zhr'-\zhr|}
\,.
\end{eqnarray}
Employment of the potential
\Eq{eqn190507n02}
instead of the potential $V^\txt{F}_\txt{cont}=-\alpha (Z-1)/r$ yields a correction to the scattering amplitude.
This correction is scaled with $Z$ as $1/Z^2$.

\vspace{0.25cm}  
In \cite{benis2004pra69-052718} 
an experimental-theoretical investigation of resonant electron scattering in collisions of hydrogen-like ions B$^{4+}$(1s) of boron 
with H$_2$ targets was reported. In particular, 
in \cite{benis2004pra69-052718} 
results of non-relativistic calculations within the R-matrix approach were presented. In order to make a certain test of our method we performed calculations for the same scattering system.
In Fig.~\ref{figure07} the differential cross section of elastic electron scattering on B$^{4+}(1s)$ is presented in the rest frame of the ion.
The solid black curve shows our results obtained by using 
the potential $V^\txt{F}_\txt{cont}=-\alpha (Z-1)/r$,
the dashed red curve displays the results calculated 
with the potential \Eq{eqn190507n02}.
By comparing them one can conclude that 
the higher orders of the perturbation theory corresponding to the interaction \Eq{coulomb-screening}
give quite a small correction in the case of boron ions
and, hence, can be neglected for heavier ions as well.

%
%

%
%

Comparing our results with those obtained using the non-relativistic R-matrix approach of \cite{benis2004pra69-052718} we see that on overall  
there is a reasonably good agreement between them. 
Nevertheless, one substantial disagreement should be mentioned:  
our results show the presence of clear resonances due to the $(2s,2p_{1/2})_{1}$ and $(2p_{3/2}^2)_{2}$ autoionizing states which are absent in the calculation \cite{benis2004pra69-052718}.
It can be explained by the difference between relativistic and non-relativistic description of the electron states which remains noticeable even for light atomic systems. 

\begin{figure}[h]
\includegraphics[width=25pc]{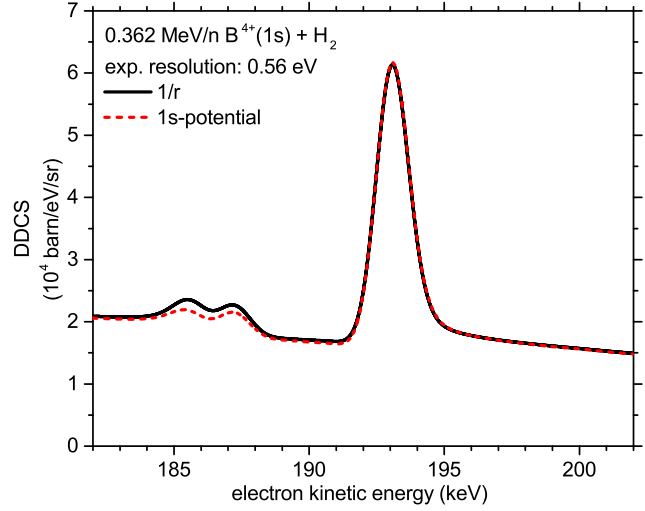}
\caption{Double differential cross section from the top panel of Fig.~\ref{figure07} convoluted with Gaussian (see Eq.~(\ref{convolutionexp})).
The solid black curves represent results of calculations with potential
$\Delta V_\txt{cont}=-\alpha/r$,
the dotted red curve corresponds to the calculation with the potential $\Delta V_\txt{cont}$
given by
\Eq{eqn190507n01}.
}
\label{figure07_2}
\end{figure}

For a comparison of our results with the experimental data of \cite{benis2004pra69-052718}, the doubly differential cross section for  collision with hydrogen (the upper panel of Fig.~\ref{figure07}) was convoluted with a Gaussian function 
\begin{eqnarray}
\left(\frac{d^2\sigma (\varepsilon_f,\theta_f)}{d\varepsilon_f d\Omega_f}\right)_{H_2,\txt{exp}}
&=&\label{convolutionexp}
\frac{1}{\sqrt{2\pi}\,\sigma_\txt{r}}\int d\varepsilon
\exp\left({\frac{-(\varepsilon-\varepsilon_f)^2}{2\sigma^2_\txt{r}}}\right) \left(\frac{d^2\sigma (\varepsilon,\theta_f)}{d\varepsilon d\Omega_f}\right)_{H_2}
\,,
\end{eqnarray}
where $\sigma_\txt{r}=0.56$ eV is the experimental resolution.
The result of the convolution is presented in Fig.~\ref{figure07_2}. 
The resonance with the $(2s)^2$ state is clearly seen in our calculations and in the calculations of \cite{benis2004pra69-052718}; however, it is not observable in the experimental data of \cite{benis2004pra69-052718}. 
Also, we note that our results obtained using \Eq{eqn190507n02}
(the dashed red line in \Fig{figure07})
are in a somewhat better agreement with the experiment.
%
%
%
%

The resonant energies of the incident electron as well as the energies and the widths of the corresponding autoionizing states for helium-like boron are presented in Table~\ref{tab5} and compared with data taken from
\cite{benis2004pra69-052718}. 
A good agreement of our results with the theoretical and experimental data of \cite{benis2004pra69-052718} shows that our approach can also be applied to relatively light systems.
%
%
\begin{table}
 \caption{\label{tab5} Same as in Table \ref{tab1} but for helium-like boron.}
\begin{tabular}{cccc}
\hline
autoionizing & $E^{'}$& $\Gamma$ & $\varepsilon_\txt{res}$ \\
state & eV& eV& eV\\
\hline
$(2s^2)_0$ &-154.5 & 0.195 & 185.8\\
$(2s^2) {}^1S$ &       & $0.194^a$ & $186.14^a$\\
\hline
$(2s2p_{1/2})_{1}$ & -153.3 & 0.008 & 187.0\\
$(2s2p_{1/2})_{0}$ & -152.9 & 0.008 & 187.4\\
$(2s2p_{3/2})_{2}$ & -152.8 & 0.008 & 187.4\\
$(2s2p) {}^3P$ &        & $0.015^a$ & $187.44^a$\\
\hline
$(2p_{1/2}^2)_0$ & -149.5 & $<$0.001 & 190.8\\
$(2p_{3/2}^2)_{2}$ & -149.4 & 0.002 & 190.9\\
$(2p_{1/2}2p_{3/2})_{1}$ & -148.9 & $<$0.001 & 191.3\\
\hline
$(2p_{1/2}2p_{3/2})_{2}$ & -147.2 & 0.192 & 193.0\\
$(2p^2) {}^1D$ &  & $0.188^a$ & $193.89^a$\\
\hline
$(2s2p_{3/2})_{1}$ & -146.5 & 0.103 & 193.8\\
$(2s2p) {}^1P$ &  & $0.088^a$ & $194.20^a$\\
\hline
$(2p_{3/2}^2)_{0}$ & -139.7 & 0.007 & 200.6\\
$(2p^2) {}^1S$ &  & $0.010^a$ & $200.90^a$\\
\hline
\multicolumn{2}{l}{$^{a}$ \cite{benis2004pra69-052718}}\\
\end{tabular}
\end{table}
%
%

\section{Conclusion}
We have considered elastic scattering of an electron on a hydrogen-like highly charged ion. The focus of the study has been on electron impact energies where autoionizing states of the corresponding helium-like ion may play an important role in the process. Compared to electron scattering on light ions the main difference in the present case is a strong field generated by the nucleus of the highly charged ions which makes it necessary to take into account the relativistic and QED effects. To this end we have developed {\it ab initio} relativistic QED theory for elastic electron scattering on hydrogen-like highly charged ions which describes in a unified and self-consistent way both the direct (Coulomb) scattering and resonant scattering proceeding via formation and consequent decay of autoionizing states.

Using this theory we have calculated scattering cross sections for a number of collision systems ranging from relatively light to very heavy ones.
As one could expect, with increasing the charge of the ionic nucleus the role of the resonant scattering decreases. However, even for ions with $Z \approx 50$ the resonances in the cross section remain clearly visible for backward scattering.

Although the presented theory has been developed first of all for the description of collisions with highly charged ions, its application for such a light system as ($e^{-}+$B$^{4+}$) demonstrates that it can be successfully used for an accurate description of a very broad range of colliding systems.

\acknowledgments
The work of D.M.V. and O.Y.A. on the calculation of the differential cross sections was supported solely by the Russian Science Foundation under Grant 17-12-01035.
The work of O.Y.A. was partly supported by Ministry of Education and Science of the Russian Federation under Grant No. 3.1463.2017/4.6.
A. B. V. acknowledges the support from the Deutsche Forschungsgemeinschaft 
(DFG, German Research Foundation) 
under Grant No 349581371 (VO 1278/4 - 1).

\appendix
\section{Coulomb scattering}
\label{appendixa}
%
The Coulomb scattering amplitude can be calculated as 
the scalar product of the in- and out-states ($\psi^{(-)}$ and $\psi^{(+)}$)
\begin{eqnarray}
S_{if}
&=&
\langle\psi^{(-)}_{\zhp_f \mu_f}|\psi^{(+)}_{\zhp_i\mu_i}\rangle
\,=\,
\int d^3 \zhr \,\psi^{(-)+}_{\zhp_f \mu_f}(\zhr)\psi^{(+)}_{\zhp_i\mu_i}(\zhr)
\,.
\end{eqnarray}
We assume that the z-axis is directed along the electron 
momentum $\zhp_i$ in the initial state.
Presenting the in- and out-states in the form
given by \Eq{eqn181017n02} 
the $S$-matrix element reads
\begin{eqnarray}
S_{if}
&=&\nonumber
N^2\sum_{jlm}\sum_{j'l'm'}M^{*}_{j'l'm'\mu_{f}}(\theta,\varphi)M_{jlm\mu_{i}}(0,0)
e^{i(\phi_{jl}+\phi_{j'l'})}i^{l-l'}
\\
&&\nonumber
\times\int d^3\zhr\, \psi^{+}_{\varepsilon'j'l'm'}(\zhr)\psi_{\varepsilon jlm}(\zhr)\\
&=&
N^2\delta(\varepsilon-\varepsilon')\sum_{jlm}
M^{*}_{jlm\mu_{f}}(\theta,\varphi)M_{jlm\mu_{i}}(0,0)e^{2i\phi_{jl}}
\,,
\end{eqnarray}
where we introduced
\begin{eqnarray}
M_{jlm\mu}(\zh{\nu})
&=&\nonumber
\Omega^{+}_{jlm}(\zh{\nu}) v_{\mu}(\zh{\nu})\\
&=&\nonumber
\sum_{m_l m_s}C^{ls}_{jm}(m_l, m_s)Y^{*}_{l m_l}(\theta,\varphi)
\chi^{+}_{m_s}v_{\mu}(\theta,\varphi)\\
&=&
\sum_{m_l m_s}C^{ls}_{jm}(m_l, m_s)Y^{*}_{l m_l}(\theta,0)
\chi^{+}_{m_s}v_{\mu}(\theta,\varphi)e^{i(m_s-m)\varphi}
\,.
\end{eqnarray}
Taking into account that 
\begin{eqnarray}
Y_{l m_l}(0,0)
&=&
\sqrt{\frac{2l+1}{4\pi}}\delta_{m_l 0}
\\
\chi^{+}_{m_s}v_\mu(0,0)&=&\delta_{\mu m_s}
\end{eqnarray}
we obtain 
\begin{eqnarray}
M_{jlm\mu}(0,0)
&=&
C^{ls}_{jm}(0,\mu)\sqrt{\frac{2l+1}{4\pi}}\sim\delta_{m\mu}
\,.
\end{eqnarray}
It is convenient to introduce the quantity 
\begin{eqnarray}
(v_{\mu})_{m_s} &=& \chi^{+}_{m_s}v_{\mu}(\theta,\varphi)
\,.
\end{eqnarray}
Then the $S$-matrix element can be written as
\begin{eqnarray}
S_{if}
&=&\nonumber
N^2\delta(\varepsilon-\varepsilon')
\sum_{jl}\sum_{m_l m_s}C^{ls}_{j\mu_i}(m_l, m_s)Y_{l m_l}(\theta,0)
[\chi^{+}_{m_s}v_{\mu_f}(\theta,\varphi)]^{*}e^{i(\mu_i-m_s)\varphi}
C^{ls}_{j\mu_i}(0,\mu_i)\sqrt{\frac{2l+1}{4\pi}}e^{2i\phi_{jl}}
\\
&=&\nonumber
(-2\pi i)\delta(\varepsilon-\varepsilon')
\\
&&
\times
N^2\frac{i}{2\pi}\sum_{m} (v_{\mu_f})^{*}_m
\,e^{i(\mu_i-m)\varphi}\sum_{jl}\sqrt{\frac{2l+1}{4\pi}}
C^{ls}_{j\mu_i}(\mu_i-m , m)C^{ls}_{j\mu_i}(0,\mu_i)e^{2i\phi_{jl}}Y_{l \mu_i-m}(\theta,0)
\,.
\end{eqnarray}

Making use of \Eq{eqn181026n001}
we get 
\begin{eqnarray}
R^{\txt{Coul}}_{if}
&=&
N^2
\frac{(-1)p}{(2\pi)^2}
\sum_{m}(v_{\mu_f})_m^{*} M_{m\mu_i}
\,,
\end{eqnarray}
where, following
\cite{burke2011b}, 
we introduced the matrix $M_{m\mu_i}$
\begin{eqnarray}
M_{m\mu_i}
&=&
4\pi
\frac{1}{2p i}
e^{i(\mu_i-m)\varphi}\sum_{jl}\sqrt{\frac{2l+1}{4\pi}}
C^{ls}_{j\mu_i}(\mu_i-m , m)C^{ls}_{j\mu_i}(0,\mu_i)Y_{l \mu_i-m}(\theta,0)e^{2i\phi_{jl}}
\,.
\end{eqnarray}
In particular, 
\begin{eqnarray}
M_{m\mu}
&=&
4\pi
\frac{1}{2p i}
\sum_{jl}\sqrt{\frac{2l+1}{4\pi}}
C^{ls}_{j\mu}(0 ,\mu)C^{ls}_{j\mu}(0,\mu)Y_{l0}(\theta,0)e^{2i\phi_{jl}}
\,,\qquad \mbox{if}\qquad m\,=\,\mu
\,,
\\
M_{m\mu}
&=&
4\pi
\frac{1}{2p i}
e^{2i\mu\varphi}\sum_{jl}\sqrt{\frac{2l+1}{4\pi}}
C^{ls}_{j\mu}(2\mu , \bar{\mu})C^{ls}_{j\mu}(0,\mu)Y_{l 2\mu}(\theta,0)e^{2i\phi_{jl}}
\,,\qquad \mbox{if}\qquad m\,=\,-\mu
\,.
\end{eqnarray}
Accordingly, we obtain
\begin{align}
M_{+1/2,+1/2}(\theta,\varphi)&=f(\theta),&M_{+1/2,-1/2}(\theta,\varphi)&=g(\theta)e^{-i\varphi}\nonumber\\
M_{-1/2,+1/2}(\theta,\varphi)&=-g(\theta)e^{i\varphi},&M_{-1/2,-1/2}(\theta,\varphi)&=f(\theta)
\,.
\end{align}

Taking into account that 
\begin{eqnarray}
Y_{l1}(\theta,0)
&=&
-Y_{l,-1}(\theta, 0)
\,=\,
-\sqrt{\frac{(2l+1)(l-1)!}{4\pi(l+1)!}}P^1_l(\cos\theta)
\\
Y_{l0}(\theta,0)
&=&
\sqrt{\frac{2l+1}{4\pi}}P_l(\cos\theta)
\\
(C^{l1/2}_{jm}(0,m))^2
&=&
\frac{|\varkappa|}{2l+1}
\\
C^{ls}_{jm}(2m , \bar{m})C^{ls}_{jm}(0,m)
&=&
\frac{1}{2l+1}\sqrt{l(l+1)}\,,\qquad j=l+\dfrac{1}{2}
\\
C^{ls}_{jm}(2m , \bar{m})C^{ls}_{jm}(0,m)
&=&
-\frac{1}{2l+1}\sqrt{l(l+1)}\,,\qquad j=l-\dfrac{1}{2}
\end{eqnarray}
we obtain 
\begin{eqnarray}
f(\theta)
&=&\label{eqn181026n11}
\dfrac{1}{2p i}\sum_{jl}|\varkappa|(e^{2i\phi_\varkappa}-1)P_l(\cos\theta)
\\
g(\theta)
&=&\label{eqn181026n12}
\dfrac{1}{2p i}\sum_{l}(e^{2i\phi_{\varkappa=-l-1}}-e^{2i\phi_{\varkappa=l}})P^1_l(\cos\theta)
\,,
\end{eqnarray}
where $\phi_{\varkappa}\equiv \phi_{jl}$ and $\varkappa$ is the Dirac quantum number
\begin{eqnarray}
\varkappa
&=&
\left(j+\frac{1}{2}\right)(-1)^{j+l+1/2}
\,.
\end{eqnarray}

Using \Eq{eqn181026n10}
the differential cross section for the Coulomb scattering 
is obtained to be 
\begin{eqnarray}
d\sigma_{if}
&=&
2\pi |R_{if}|^2
\frac{\varepsilon}{p}
\delta(\mee_i-\mee_f)
\frac{d^3\zhp}{(2\pi)^3}
\,,
\\
\frac{d\sigma_{if}}{d\zhnu}
&=&
\left|\sum_{m}(v_{\mu_f})_m^{*} M_{m\mu_i}\right|^2
\,.
\end{eqnarray}

\section{Numerical calculation of the Coulomb amplitudes}
\label{appendixb}
The Coulomb amplitudes are given by series
Eqs.~\Br{eqn181026n11}, \Br{eqn181026n12}.
These series are not convenient for a direct numerical calculation.
However, the leading part of these series can be calculated analytically 
and the remaining part can be easily summed up numerically.

The Coulomb phase shifts for the potential $V=-\alpha Z/r$ read 
\begin{eqnarray}
\phi_{\varkappa}
&=&\label{coulomb-phase}
\arg\Gamma(\gamma-i\nu) + \eta -\frac{1}{2}\pi\gamma+ \frac{\pi}{2}(l+1)
\,,
\end{eqnarray}
where
\begin{eqnarray}
e^{2i\eta}
&=&
\frac{-\varkappa+i\nu'}{\gamma+i\nu}
\\
\nu
&=&
\frac{\varepsilon}{p}\alpha Z
\\
\nu'
&=&
\frac{m_e}{p}\alpha Z
\\
\gamma
&=&
\sqrt{\varkappa^2-\alpha^2 Z^2}
\,,
\end{eqnarray}
where $\alpha$ is the fine-structure constant and 
$m_e$ is the electron mass.

Following \cite{johnson-scatter} 
we introduce
\begin{eqnarray}
a_{\varkappa}(\gamma)
&=&\nonumber
e^{2i\phi_{\varkappa}}-1
\\
&=&
(-1)^{l+1}
\left(
\frac{-\varkappa+i\nu'}{\gamma+i\nu}
\right)
\frac{\Gamma(\gamma-i\nu)}{\Gamma(\gamma+i\nu)}
e^{-i\pi\gamma} -1
\,.
\end{eqnarray}
Then the amplitudes \Br{eqn181026n11} and \Br{eqn181026n12} 
can be written as  
\begin{eqnarray} 
f(\theta)
&=&\nonumber
\frac{1}{2ip}
\sum\limits_{jl}
|\varkappa|a_{\varkappa}
P_l(\cos\theta)
\\
&=&
\frac{1}{2ip}
\sum\limits_{l}
\left[
(l+1)a_{\varkappa=-l-1}
+
l a_{\varkappa=l}
\right]
P_l(\cos\theta)
\\
g(\theta)
&=&
\frac{1}{2ip}
\sum\limits_{l}
\left(
a_{\varkappa=-l-1}
-
a_{\varkappa=l}
\right)
P^1_l(\cos\theta)
\,.
\end{eqnarray}

Now we introduce the approximate amplitudes 
\begin{eqnarray}
\tilde{f}(\theta)
&=&\label{eqn181030n03}
\frac{1}{2ip}
\sum\limits_{l}
\left[
(l+1)\tilde{a}_{\varkappa=-l-1}
+
l\tilde{a}_{\varkappa=l}
\right]
P_l(\cos\theta)
\\
\tilde{g}(\theta)
&=&\label{eqn181030n04}
\frac{1}{2ip}
\sum\limits_{l}
\left[
\tilde{a}_{\varkappa=-l-1}
-
\tilde{a}_{\varkappa=l}
\right]
P^1_l(\cos\theta)
\,,
\end{eqnarray}
where 
\begin{eqnarray}
\tilde{a}_{\varkappa}
&=&
a_{\varkappa}(\gamma=|\varkappa|)
\end{eqnarray}
and, correspondingly, 
\begin{eqnarray}
\tilde{a}_{\varkappa=-l-1}
&=&
(l+1+i\nu')
\frac{\Gamma(l+1-i\nu)}{\Gamma(l+2+i\nu)}
-1
\\
\tilde{a}_{\varkappa=l}
&=&
(l-i\nu')
\frac{\Gamma(l-i\nu)}{\Gamma(l+1+i\nu)}-1
\,.
\end{eqnarray}
The series
Eqs.~\Br{eqn181030n03}, \Br{eqn181030n04}
can be summed up analytically \cite{johnson-scatter} 
\begin{eqnarray}
\tilde{f}(\theta)
&=&
\frac{\Gamma(1-i\nu)}{\Gamma(1+i\nu)}
e^{i\nu\ln\sin^2(\theta/2)}
\left[
\frac{\nu}{2p}\csc^2(\theta/2)+\frac{\nu'-\nu}{2p}
\right]
\\
\tilde{g}(\theta)
&=&
\frac{\Gamma(1-i\nu)}{\Gamma(1+i\nu)}
e^{i\nu\ln\sin^2(\theta/2)}
\left[
\frac{\nu'-\nu}{2p} \cot(\theta/2)
\right]
\,.
\end{eqnarray}

As a result, the Coulomb amplitudes can be rewritten as 
\begin{eqnarray}
f(\theta)
&=&\nonumber
\tilde{f}(\theta)
+
\frac{1}{2ip}
\sum\limits_{jl}
|\varkappa|(a_{\varkappa}-\tilde{a}_{\varkappa})
P_l(\cos\theta)
\\
&=&
\tilde{f}(\theta)
+
\frac{1}{2ip}
\sum\limits_{l}
\left[
(l+1)(a_{\varkappa=-l-1}-\tilde{a}_{\varkappa=-l-1})
+
l (a_{\varkappa=l}-\tilde{a}_{\varkappa=l})
\right]
P_l(\cos\theta)
\\
g(\theta)
&=&
\tilde{g}(\theta)
+
\frac{1}{2ip}
\sum\limits_{l}
\left(
a_{\varkappa=-l-1}-\tilde{a}_{\varkappa=-l-1}
-
a_{\varkappa=l}+\tilde{a}_{\varkappa=l}
\right)
P^1_l(\cos\theta)
\,, 
\end{eqnarray}
where the corresponding series can be easily calculated numerically.

%
%
%

%
%

\end{document}